\documentclass{aa}  
\usepackage{graphicx}
\usepackage[version=4]{mhchem}
\usepackage{units}
\usepackage{amsmath,amssymb}
\usepackage[utf8]{inputenc}
\usepackage[T1]{fontenc}
\usepackage{xcolor}
\usepackage{appendix}
\usepackage{booktabs}
\usepackage{multirow}
\DeclareUnicodeCharacter{2212}{-}
\usepackage{graphicx}
\usepackage{soul}
\usepackage{txfonts}
\begin{document}

   \title{\ce{H2S} ice sublimation dynamics}

   \subtitle{Experimentally constrained binding energies, entrapment efficiencies, and snowlines}

   \author{Julia C. Santos
          \inst{1}
          \and
          Elettra L. Piacentino\inst{2}
          \and
          Jennifer B. Bergner\inst{3}
          \and
          Mahesh Rajappan\inst{2}
          \and
          Karin I. \"{O}berg\inst{2}
          }


            \institute{Laboratory for Astrophysics, Leiden Observatory, Leiden University, PO Box 9513, 2300     RA Leiden, The Netherlands\\
                \email{santos@strw.leidenuniv.nl}
         \and Center for Astrophysics, Harvard \& Smithsonian, 60 Garden St., Cambridge, MA 02138, USA
         \and UC Berkeley Department of Chemistry, Berkeley, CA 94720, USA
                }


 
  \abstract
   {Hydrogen sulfide (\ce{H2S}) is thought to be an important sulfur reservoir in interstellar ices. It serves as a key precursor to complex sulfur-bearing organics, and has been proposed to play a significant role in the origin of life. Although models and observations both suggest \ce{H2S} to be present in ices in non-negligible amounts, its sublimation dynamics remain poorly constrained.} 
   {In this work, we present a comprehensive experimental characterization of the sublimation behavior of \ce{H2S} ice under astrophysically-relevant conditions.}
   {We used an ultrahigh vacuum chamber to deposit pure multilayer \ce{H2S} ice, submonolayer \ce{H2S} ice on top of compact amorphous solid water (cASW), as well as ice mixtures of \ce{H2S} and \ce{H2O}. The sublimation behavior of \ce{H2S} was monitored with a quadrupole mass spectrometer (QMS) during temperature-programmed desorption (TPD) experiments. These experiments are used to determine binding energies and entrapment efficiencies of \ce{H2S}, which are then employed to estimate its snowline positions in a protoplanetary disk midplane.}
   {We derive mean binding energies of $3159\pm46$ K for pure \ce{H2S} ice and $3392\pm56$ K for submonolayer \ce{H2S} desorbing from a cASW surface. These values correspond to sublimation temperatures of around 64 K and 69 K in the disk midplane, placing its sublimation fronts at radii just interior to the \ce{CO2} snowline. We also investigate the entrapment of \ce{H2S} in water ice and find it to be highly efficient, with $\sim75-85\%$ of \ce{H2S} remaining trapped past its sublimation temperature for \ce{H2O}:\ce{H2S} mixing ratios of $\sim$5$-$17:1. We discuss potential mechanisms behind this efficient entrapment.}
   {Our findings imply that, in protoplanetary disks, \ce{H2S} will mostly be retained in the ice phase until water crystallizes, at radii near the water snowline, if it forms mixed into water ice. This has significant implications for the possibility of \ce{H2S} being incorporated into icy planetesimals and its potential delivery to terrestrial planets, which we discuss in detail.}


   \maketitle
%

\section{Introduction} \label{sec:intro}

The interchange between solids and gas plays a major role in the chemical composition and structure of star- and planet-forming regions \citep{Bergin1997, Aikawa2002, Viti2004, Henning2013, He2016a, Oberg2023}. At the cold temperatures of interstellar clouds (typically $\sim$10$-$20 K), most molecules adsorb onto dust grains, forming ice mantles that undergo rich solid-state chemical processes. The result is a lavish icy chemical reservoir---spanning from simple molecules such as \ce{CO} and \ce{H2O} to complex organics \citep{Herbst2009, vanDishoeck2014, Linnartz2015, Oberg2016, Cuppen2024}. As this interstellar material collapses into an emerging young stellar object, increasing temperatures caused by the heat from the protostar enable the diffusion of ice species---further facilitating chemical reactions (see, e.g., \citealt{Cuppen2017})---and eventually lead to their thermal sublimation. The locations of these sublimation fronts, or snowlines, determine the physical state in which specific molecules are available for incorporation into forming planets and planetesimals, thus shaping their solid and atmospheric constitution \citep{Oberg2011a, Henning2013, Madhusudhan2019}. Such locations are dictated by a combination of desorption kinetics, set by a species' binding energies, and the efficiencies with which it is trapped within less volatile ice matrices. It is thus paramount to characterize the thermal sublimation behavior of ice species in order to understand the chemical evolution of environments where young solar system bodies are forming.

One particularly riveting volatile interstellar molecule is hydrogen sulfide (\ce{H2S}). Following its first detection by \cite{Thaddeus1972} in seven Galactic sources, it has since been observed in the gas phase towards a range of interstellar and protoplanetary environments: from clouds \citep{Minh1989, Neufeld2015} to dense cores and protostars \citep{Minh1990, vanDishoeck1995, Hatchell1998, Vastel2003, Wakelam2004} to protoplanetary disks \citep{Phuong2018, Riviere-Marichalar2021, Riviere-Marichalar2022}. On the other hand, interstellar \ce{H2S} ice has yet to be observed, with abundance upper limits in prestellar cores and protostellar envelopes estimated as $\lesssim$1\% with respect to \ce{H2O} \citep{Smith1991, Jimenez-Escobar2011, McClure2023}. This non-detection is likely associated with the intrinsic limitations of astronomical observations in the solid state (e.g., the broadness of the ice features and their high degeneracy). Indeed, the strongest IR feature of \ce{H2S} ice (its S-H stretching modes at 3.93 $\mu$m) is particularly challenging to unequivocally observe due to its broad profile and because it overlaps both with \ce{CH3OH} combination modes (a major ice component) and with the S-H stretching modes of simple thiols \citep{Jimenez-Escobar2011, Hudson2018}. While the latter are less concerning due to the low expected abundances of thiols, both factors still complicate confidently assigning absorption features in this region to \ce{H2S}. Nevertheless, \ce{H2S} is predicted by chemical models to be very efficiently formed in ices via the successive hydrogenation of \ce{S} atoms (see, e.g., \citealt{Garrod2007, Druard2012, Esplugues2014, Vidal2017, Vidal2018}):

\begin{equation}
    \ce{S} \xrightarrow{\text{+ H}} \ce{HS} \xrightarrow{\text{+ H}} \ce{H2S}.
\end{equation} 

\noindent Indeed, it has been observed to be a major sulfur carrier in the comae of comets \citep{Mumma2011, LeRoy2015, Biver2015, Calmonte2016}, which are thought to (at least partially) inherit the ice material from the prior pre- and protostellar evolutionary stages (e.g., \citealt{Bockelee-Morvan2000, Altwegg2017, Rubin2018, Drozdovskaya2019}). Cometary \ce{H2S} abundances relative to \ce{H2O} range between $\sim$0.13$-$1.75\% (\citealt{Calmonte2016} and references therein), with measurements from the coma of comet 67P/Churymov-Gerasimenko (hereafter 67P) by the Rosetta Orbiter Sensors for Ion and Neutral Analysis (ROSINA) instrument on board of the Rosetta spacecraft yielding \ce{H2S}/\ce{H2O} abundances of $1.06\pm0.05 \%$ \citep{Calmonte2016}. These findings point to \ce{H2S} as the main volatile sulfur carrier in 67P. Consequently, while the upper limits on \ce{H2S} ice are sufficient to rule it out as the main interstellar sulfur reservoir (see, e.g., \citealt{Jimenez-Escobar2011}), cometary inheritance from interstellar ices remains a plausible hypothesis within current observational constraints. Moreover, observed gas-phase \ce{H2S} abundances toward solar-mass protostars---attributed to the sublimation of ices in the hot core region---suggest \ce{H2S} to be an important gaseous sulfur carrier \citep{Drozdovskaya2018}. Given that gas-phase routes to \ce{H2S} cannot account for its detected gaseous abundances, all evidence point to it being present in interstellar ices at a level of $\sim$1\% with respect to \ce{H2O}.


Irrespective of its physical state, \ce{H2S} can serve as a important source of sulfur during the chemical evolution of star- and planet-forming regions. As a solid, it has been shown both experimentally and by chemical models to initiate a prolific sulfur network by producing \ce{HS} radicals and \ce{S} atoms---induced either by energetic processing or interactions with H atoms---that readily react with other ice species \citep{Moore2007, Ferrante2008, Garozzo2010, Jimenez-Escobar2014, Chen2015, Laas2019, Santos2024a, Santos2024c}. Solid \ce{H2S} and its reaction products might then be incorporated into icy planetesimals, which in turn might deliver them to terrestrial planets during events such as our Solar System's late heavy bombardments. This is particularly relevant to theories on the origins of life, as \ce{H2S} has been proposed as a key energy source for early metabolic pathways predating oxygenic photosynthesis \citep{Olson2016}. More broadly, sulfur-bearing compounds have long been recognized as fundamental to biological systems. Additionally, \ce{H2S} can also undergo solid-state acid-base reactions with \ce{NH3} to form ammonium hydrosulfide (\ce{NH4^+SH^-}) at temperatures as low as 10 K \citep{Loeffler2015, Vitorino2024, Slavicinska2024b}. This salt has been detected at very high abundances in the grains of comet 67P by the ROSINA instrument \citep{Altwegg2022}, and is proposed to be a major carrier of the 6.85 $\mu$m band assigned to \ce{NH4^+} in ices, as well as a significant sink to the conspicuous missing sulfur problem \citep{Slavicinska2024b}. As a gas, \ce{H2S} can contribute to the elemental abundance of sulfur in planetary atmospheres---which in turn might help tracing the planet's formation history \citep{Oberg2011a, Polman2023, Tsai2023}.


Despite its pivotal role in the sulfur network of star- and planet-forming regions, a comprehensive characterization of the thermal sublimation behavior of \ce{H2S} is still lacking from the literature, with no experimentally-determined binding energies or entrapment efficiencies available to date. We aim to bridge this gap with this work. The paper is outlined as follows: in Section \ref{sec:methods}, the experimental setup and procedures are described. In section \ref{sec:res_dis}, we report and discuss our results, including experimentally-derived binding energies as well as entrapment efficiencies in \ce{H2O} ice. The corresponding locations of the \ce{H2S} sublimation fronts and their astrophysical implications are discussed in Section \ref{sec:astro}. Finally, in Section \ref{sec:concl} we summarize our main findings.

\section{Methods} \label{sec:methods}
\subsection{The setup}

This work utilized the experimental setup SPACE-KITTEN\footnote{Surface Processing Apparatus for Chemical Experimentation–Kinetics of Ice Transformation in Thermal ENvironments}, which has been described in detail elsewhere \citep{Simon2019, Simon2023}. Briefly, the setup consists of an ultrahigh vacuum (UHV) chamber with base pressure at room temperature of $\sim$4$\times$10$^{-9}$ Torr. At the center of the chamber, an IR-transparent CsI window is mounted on a sample holder attached to a closed-cycle helium cryostat with a DMX-20B interface (which decouples the sample holder from the cryostat's cold tip, thus preventing vibrations). The substrate temperature can be varied between 12 and 300 K using a resistive thermofoil heater, and is monitored by two silicon diode sensors with a precision of $\pm$0.1 K and absolute accuracy of $\sim$2 K. \ce{H2S} (Sigma-Aldrich, purity $\geq$99.5\%) and \ce{H2O} are admitted into the chamber, either individually or as a mixture, through a stainless steel tube doser with a diameter of 4.8 mm at normal incidence to the substrate. The doser outlet is located at 1 inch from the substrate during deposition and the base pressure of the gas line is $\sim$10$^{-4}$ Torr. The \ce{H2O} sample was prepared by purifying deionized water through multiple freeze-pump-thaw cycles in a liquid nitrogen bath. Moreover, \ce{^{13}CO2} (Sigma-Aldrich, purity $99\%$; $<$3 atom \% \ce{^{18}O}, 99 atom \% \ce{^{13}C}) is also utilized for a control experiment mixed with \ce{H2O}. The minor isotopologue is chosen to avoid any potential residual atmospheric contamination from interfering with the analysis. Ice growth is monitored by a Bruker Optics Vertex 70 Fourier-transform infrared spectrometer (FTIR) in transmission mode, while the gas-phase composition in the chamber is sampled by a Pfeiffer Vaccum Inc. PrismaPlus QMG 220 quadrupole mass spectrometer (QMS).

\begin{table*}[htb!]
\centering
\caption{List of experiments.}
\label{tab:exp_list} 
\begin{tabular}{lcccc}  
\toprule\midrule
Experiment$^{a}$                          &   X coverage (ML)              &   \ce{H2O} coverage (ML)           &   Total mixture coverage (ML)   &   \ce{H2O}:X \\
\midrule
\multicolumn{5}{c}{X = \ce{H2S}}\\
\midrule
\ce{H2S}                            &    90                          &    $-$                             &    $-$                          &   $-$\\
\ce{H2S}                            &    73                          &    $-$                             &    $-$                          &   $-$\\  
\ce{H2S}                            &    45                          &    $-$                             &    $-$                          &   $-$\\  
\ce{H2S}                            &    32                          &    $-$                             &    $-$                          &   $-$\\  
\ce{H2S}                            &    19                          &    $-$                             &    $-$                          &   $-$\\
                                    &                                &                                    &                                 &\\
\ce{H2O}$\to$\ce{H2S}               &    1.6                         &    42                              &    $-$                          &   $-$\\
\ce{H2O}$\to$\ce{H2S}               &    1.2                         &    40                              &    $-$                          &   $-$\\
\ce{H2O}$\to$\ce{H2S}               &    0.6                         &    40                              &    $-$                          &   $-$\\
\ce{H2O}$\to$\ce{H2S}               &    0.5                         &    39                              &    $-$                          &   $-$\\
\ce{H2O}$\to$\ce{H2S}               &    0.3                         &    39                              &    $-$                          &   $-$\\
\ce{H2O}$\to$\ce{H2S}               &    0.2                         &    47                              &    $-$                          &   $-$\\
                                    &                                &                                    &                                 &\\
\ce{H2O}:\ce{H2S}                   &    7.0                         &    36                              &    43.0                         &   5.1:1\\
\ce{H2O}:\ce{H2S}                   &    5.1                         &    38                              &    43.1                         &   7.5:1\\
\ce{H2O}:\ce{H2S}                   &    4.2                         &    39                              &    43.2                         &   9.3:1\\
\ce{H2O}:\ce{H2S}                   &    2.3                         &    40                              &    42.3                         &   17:1\\
\midrule
\multicolumn{5}{c}{X = \ce{^{13}CO2}}\\
\midrule
\ce{H2O}:\ce{^{13}CO2}              &    2.6                         &    37                              &    39.6                         &   14:1\\
\bottomrule
\multicolumn{5}{l}{\footnotesize{$^a$ Arrows (\ce{H2O}$\to$X) denote sequential depositions (X on top of \ce{H2O}); colons (\ce{H2O}:X) denote ice mixtures of \ce{H2O} and X.}}\\
\end{tabular}
\end{table*}

\subsection{The experiments}

The experiments performed in this work are summarized in Table \ref{tab:exp_list}. We derive binding energies of \ce{H2S} in two scenarios: for pure multilayer \ce{H2S} ice and for submonolayer \ce{H2S} on top of multilayer compact amorphous solid water (cASW). The latter was chosen as a substrate because it is generally thought to be more representative of interstellar water ice than porous counterparts (e.g., \citealt{Acolla2013}). In order to grow cASW, the substrate is kept at 100 K during the \ce{H2O} ice deposition \citep{Bossa2012}, and is subsequently cooled down to 15 K before depositing \ce{H2S}. The pure multilayer \ce{H2S} ices are deposited at 15 K directly. Entrapment experiments are performed for water-dominated ice mixtures of \ce{H2S} or \ce{^{13}CO2} deposited at 15 K. Prior to dosing, the mixtures are prepared in a gas manifold within one hour of the experiment. In all cases, after the deposition is completed, temperature programmed desorption (TPD) experiments are performed by heating up the substrate temperature linearly at a rate of 2 K min$^{-1}$. The desorbed species are immediately ionized by 70 eV electron impact and are monitored by the QMS. Their desorption rates are then used to derive the binding energies and entrapment efficiencies. Any contamination from background \ce{H2O} deposition is negligible. Based on the infrared absorption features of \ce{H2O} observed after \ce{H2S} deposition in the pure ice experiments, the background \ce{H2O} deposition rate is at most 0.1 ML/min. This is at least one order of magnitude lower than the \ce{H2S} deposition rate in the multilayer experiments and a factor of $\sim$5 lower in the submonolayer experiments. Even in the latter case, we find no evidence of significant \ce{H2O} codeposition with \ce{H2S}, as no additional desorption feature is observed with the QMS following the submoloayer desorption of \ce{H2S}---which would be expected if a non-negligible amount of porous ASW were forming due to background deposition at 15 K.

The surface coverage of the ice species is quantified by two approaches. For ices with thicknesses $\gtrsim$1 ML, taken as the typical approximation of 1 ML = 10$^{15}$ molecules cm$^{-2}$, their infrared absorbance bands are reliably detected above the instrumental limit of the spectrometer. In such cases, the IR integrated absorbance ($\int Abs(\tilde{\nu})d\nu$) of the species is converted to absolute abundance using a modified Beer-Lambert law:

\begin{equation}
    N_X=\ln10\frac{\int Abs(\tilde{\nu})\text{d}\tilde{\nu}}{A(X)},
    \label{eq:N_RAIRS}
\end{equation}

\noindent where $N_X$ is the species' column density in molecules cm$^{-2}$ and $A(X)$ is its absorption band strength in cm molecule$^{-1}$. We use $A(\ce{H2S})_{\text{S-H str}}$ = $1.69\times10^{-17}$ cm molecule$^{-1}$ for pure \ce{H2S} ices and $A(\ce{H2S})_{\text{S-H str}}$ = $1.66\times10^{-17}$ cm molecule$^{-1}$ for \ce{H2S} mixed with \ce{H2O}, as derived by \cite{Yarnall2022}. For \ce{H2O} and \ce{^{13}CO2}, we use $A(\ce{H2O})_{\text{O-H str}}$ = $2.2\times10^{-16}$ cm molecule$^{-1}$ and $A(\ce{^{13}CO2})_{\text{C=O str}}$ = $1.15\times10^{-16}$ cm molecule$^{-1}$, taken from \cite{Bouilloud2015} based on the values reported by \cite{Gerakines1995}. Uncertainties in the band strengths are the main source of error in the ice coverage estimation, and are assumed to be 10\% to account for possible variations caused by temperature and mixing conditions. This uncertainty is then propagated throughout the analysis. For the cases in which ices are deposited with submonolayer coverages, IR absorption bands are not or barely detected and the coverage is estimated by the species' integrated QMS signal during TPD corrected by a scaling factor (see Appendix \ref{Appendix:subML_calib} for more details).

\subsection{The analysis}
\label{sec:2_analysis}
The \ce{H2S} binding energies are derived by fitting the measured TPD curves with the Polanyi-Wigner equation:

\begin{equation}
    -\frac{d\theta}{dT}=\frac{\nu}{\beta}\theta^n e^{-E_\text{des}/T},
    \label{eq:Polanyi}
\end{equation}

\noindent where $\theta$ is the ice coverage in monolayers, $T$ is the ice temperature in K, $\beta$ is the heating rate in  K s$^{-1}$, $\nu$ is the preexponential factor, $n$ is the kinetic order, and $E_\text{des}$ is the desorption energy in K. The kinetic order is a dimentionless quantity that indicates the influence of the species' concentration on its desorption rate. In the multilayer regime, desorption is independent of the ice thickness ($n=0$), whereas in the submonolayer regime it is proportional to the ice coverage ($n=1$). This reflects the constant number of adsorbates available for desorption at any given time in the former case, as opposed to the varying number in the latter. The preexponential factor is associated with the molecule's frequency of vibration in the adsorbate$-$surface potential well, and its units depend on the kinetic order (ML$^{1-n}$ s$^{-1}$). For a given temperature $T$, we derive the preexponential factor following the transition state theory (TST) approach and approximating the partition function of the species in the adsorbed state to unity (that is, assuming it to be fully immobile; \citealt{Tait2005, Minissale2022}):

\begin{equation}
    \nu_\text{TST}=\frac{k_\text{B}T}{h}q^\ddag_\text{tr,2D}q^\ddag_\text{rot,3D}.
    \label{eq:nuTST}
\end{equation}

\noindent Where $k_\text{B}$ is the Boltzmann constant, $h$ is the Planck constant, and $q^\ddag_\text{tr,2D}$ and $q^\ddag_\text{rot,3D}$ are the transitions state's partition functions of translation and rotation, respectively. All constants used here are in the MKS unit system. The translational motion perpendicular to the surface can be neglected, resulting in a translational partition functional parallel to the surface plane:

\begin{equation}
    q^\ddag_\text{tr,2D} = \frac{A_S 2\pi m k_\text{B}T}{h^2},
    \label{eq:qTR}
\end{equation}

\noindent where $m$ is the mass of the species in kg and $A_S$ is the surface area per adsorbed molecule, fixed to the typical value of 10$^{-19}$ m$^2$. The rotational partition function is given by:

\begin{equation}
    q^\ddag_\text{rot,3D} = \frac{\sqrt{\pi}}{\sigma h^3}(8\pi^2 k_\text{B}T)^{3/2}\sqrt{I_xI_yI_z},
    \label{eq:qROT}
\end{equation}

\noindent where $\sigma$ is the species' symmetry number (i.e., the number of indistinguishable rotated positions) and $I_xI_yI_z$ is the product of its principal moments of inertia. The choice of $T$ values used to derive $\nu_\text{TST}$ will be discussed in Sections \ref{sec:BE_multiL} and \ref{sec:BE_subML} for their respective coverage regimes, while the rest of the parameters employed in equations \ref{eq:qTR} and \ref{eq:qROT} are listed in Table \ref{tab:H2S_nuTST_params}. The inertia moments of \ce{H2S} were calculated using Gaussian 16e \citep{g16} at the M06-2X/aug-cc-pVTZ level of theory \citep{Dunning1989, Zhao2008}. The resulting $\nu_\text{TST}$ are then used to fit equation \ref{eq:Polanyi} to the TPD curves and derive the corresponding binding energies.

\begin{table}[htb!]
\centering
\caption{Summary of parameters used to derive $\nu_\text{TST}$ for \ce{H2S} (see equations \ref{eq:nuTST}, \ref{eq:qTR}, \ref{eq:qROT}).}
\label{tab:H2S_nuTST_params} 
\begin{tabular}{lc}  
\toprule\midrule
Parameter                       &   Value\\
\midrule
$m$ (kg)                        &   5.66$\times$10$^{-26}$\\
$\sigma$                        &   2\\
$I_x$ (amu $\text{\AA}^2$)      &   5.81066\\  
$I_y$ (amu $\text{\AA}^2$)      &   6.71494\\ 
$I_z$ (amu $\text{\AA}^2$)      &   12.52560\\
\bottomrule
\end{tabular}
\end{table}

When a more volatile species is embedded within a less volatile ice matrix, it can remain trapped in the solid phase beyond its expected sublimation temperature. The entrapment behavior of \ce{H2S} in \ce{H2O}-dominated ices is also investigated here, with \ce{^{13}CO2} mixed into a water-rich ice included as a control experiment. As water ice crystallizes, it creates new channels that allow trapped species to reach the surface, producing the so-called ``molecular volcano'' desorption peaks \citep{Smith1997}. Entrapment efficiencies are calculated as the ratios of the integrated volcano-desorption TPD peak of the more volatile species (\ce{H2S} or \ce{^{13}CO2}) over the integrated area of its entire TPD curve. The integration bounds for the volcano peaks are defined by the temperature range of the water desorption peak, thus accounting for all volatiles sublimating concurrently with water. All mixed-ice experiments were conducted with a fixed total coverage of $\sim$40 ML to eliminate the influence of the ice thickness and isolate the effect of the mixing ratio on the entrapment efficiencies.
 
\section{Results and discussion} \label{sec:res_dis}
\subsection{Binding energies}
\label{sec:bindingEs}
\subsubsection{\ce{H2S}$-$\ce{H2S}}
\label{sec:BE_multiL}

\begin{figure}[htb!]\centering
\includegraphics[scale=0.37]{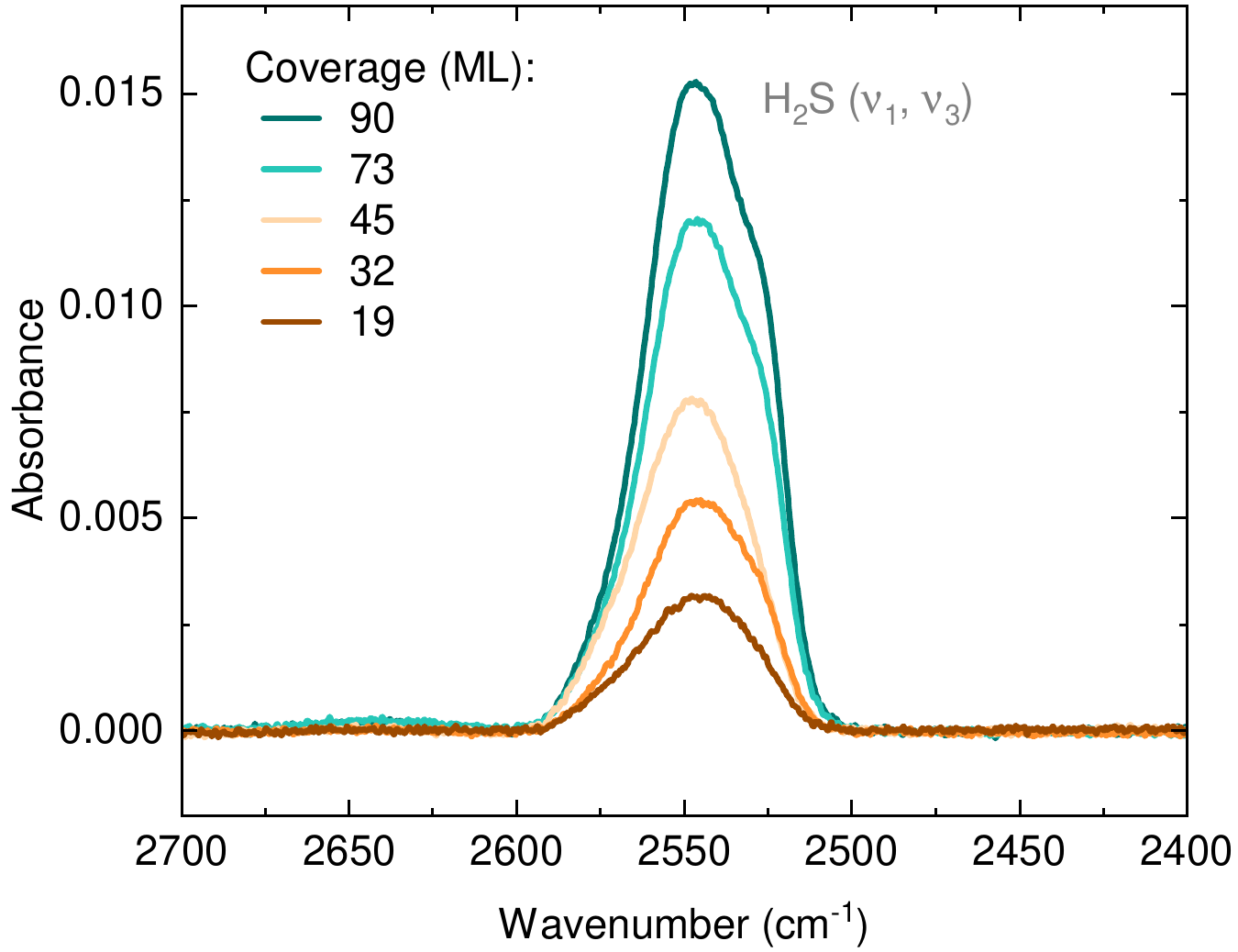}
\caption{Infrared spectra recorded after the deposition of the five different multilayer \ce{H2S} ice thicknesses on top of the CsI window. The spectra are centered on the frequency range of the S-H stretching modes of \ce{H2S} ($\nu_1$ and $\nu_3$).}
\label{fig:IR_pure_all}
\end{figure}

\begin{figure*}[htb!]\centering
\includegraphics[scale=0.3]{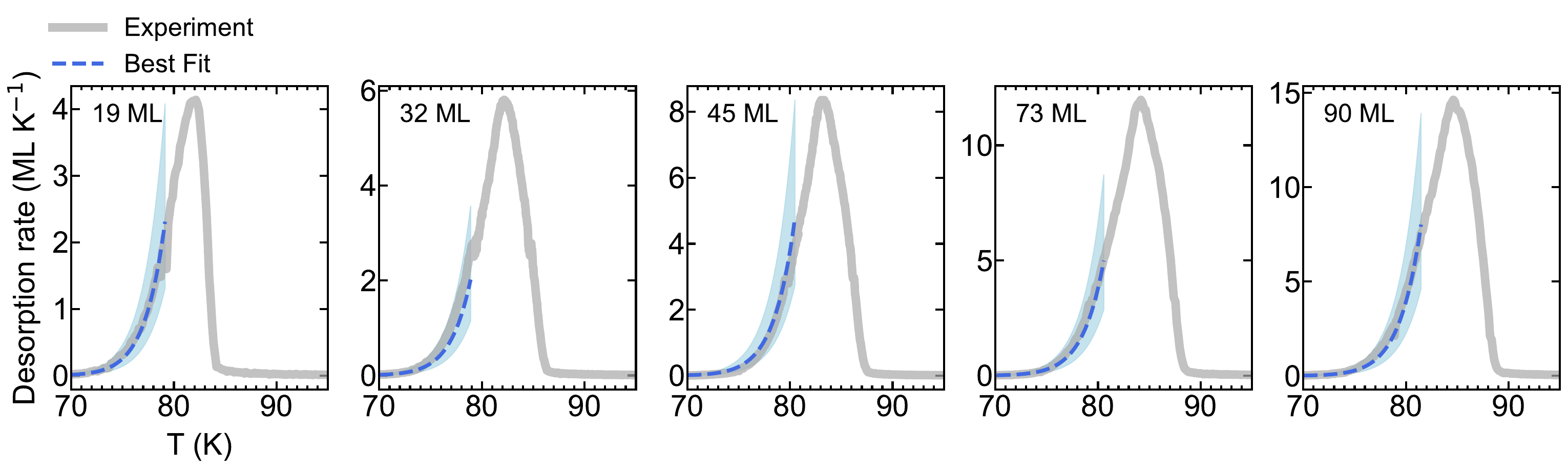}
\caption{TPD curves of the multilayer \ce{H2S} ice experiments. The experimental data is shown in gray, while the zeroth-order Polanyi-Wigner models are shown with the dashed blue curve. The shaded blue region indicates the 1$\sigma$ uncertainty.}
\label{fig:TPD_multi_leadingedge}
\end{figure*}

Figure \ref{fig:IR_pure_all} shows the IR spectra obtained after completing the deposition of the pure \ce{H2S} ice in the multilayer experiments. The figure focuses on the frequency range of the S-H stretching modes of \ce{H2S}---its most prominent IR feature. Five different thicknesses (19, 32, 45, 73, and 90 ML) were used to derive the \ce{H2S}$-$\ce{H2S} binding energy; that is, the binding energy dominated by \ce{H2S} interactions with other \ce{H2S} molecules. The TPD curves for each coverage are shown in gray in Figure \ref{fig:TPD_multi_leadingedge}, measured from the mass-to-charge (m/z) ratio of 34, which corresponds to the molecular ion of \ce{H2S} and represents its dominant mass signal in electron-impact mass spectrometry at 70 eV ionization energy\footnote{https://webbook.nist.gov}.

To fit the TPD curves with the Polanyi-Wigner equation (eq. \ref{eq:Polanyi}), we must first determine the preexponential factor $\nu_{\text{TST}}$ of \ce{H2S} using equations \ref{eq:nuTST}, \ref{eq:qTR}, and \ref{eq:qROT}. Since these equations are temperature dependent, we opt to derive $\nu_{\text{TST}}$ at the maximum temperature where the experimental curve is still well described by an exponential behavior. This is to ensure consistency in the data analysis: by choosing the maximum temperature within the exponential range, we ensure that $\nu_\text{TST}$ is consistent with the region within which the model is fitted. This temperature is determined on the basis of its adjusted R$^2$ value, a coefficient that measures the proportion of variation in outcomes explained by the model, while adjusting for the number of predictors used. This adjustment is particularly important in this context, as R$^2$ naturally increases with the temperature, even without an improvement in model fit. The maximum temperatures are 79.2, 78.9, 80.5, 80.6, and 81.5 K for the respective coverages of 19, 32, 45, 73, and 90 ML.

As the substrate is warmed up, the morphology of the \ce{H2S} ice gradually changes toward more ordered configurations, with amorphous-to-crystalline transitions beginning at temperatures as low at $\sim$30 K, and nearing completion by $\sim$60 K \citep{Fathe2006, Mifsud2024}. Although the crystallization process appears to be largely finalized by the onset of desorption ($\gtrsim$70 K), the S-H stretching modes of the \ce{H2S} ice continue to blueshift as the temperature rises beyond this threshold (see Appendix \ref{Appendix:H2S_multiL_IR_TPD} and \citealt{Mifsud2024}), indicating an ongoing reorientation of the ice until its complete desorption. This continued reorientation may contribute to deviations from zeroth-order kinetics near the desorption peak, disrupting the exponential trend. Our approach to the choice of temperature reduces the contribution from these confounding factors to the calculation of the preexponential factor.

An alternative, commonly-employed approach is to instead use the peak desorption temperature of the TPD curve. In the case of \ce{H2S}, the temperature differences between the two approaches and their overall impact on the desorption parameters are small: the $\nu_{\text{TST}}$ values derived from the peak desorption T for each coverage (respectively, 82.1, 82.2, 83.0, 84.2, and 84.6 K) deviate by $\lesssim$16\% from the values derived by our preferred method, resulting in a less than 3\% variation in the corresponding binding energies. Uncertainties in the derived $\nu_\text{TST}$ values stemming from the absolute error of 2 K in the temperature readout are of $\sim$1\%. We emphasize that the \ce{H2S}$-$\ce{H2S} binding energy derived in this work corresponds to crystalline \ce{H2S}. However, given that the crystallization of \ce{H2S} is largely complete by temperatures much lower than its onset of desorption, any thermal desorption of pure \ce{H2S} from interstellar ices is expected to occur from its crystalline phase.

The $\nu_{\text{TST}}$ values derived for each coverage are used to fit their respective TPD curves with the zeroth-order Polanyi-Wigner equation. Following the above reasoning, the fit is performed for the temperature range where the curve maintains an exponential behavior; that is, until the same maximum temperature for which $\nu_{\text{TST}}$ was calculated. We performed a Monte Carlo analysis using 10,000 independent trials to incorporate and quantify the uncertainties in the ice coverage, absolute substrate temperature, and $\nu_{\text{TST}}$. In each trial, temperature, coverage, and $\nu_{\text{TST}}$ values were randomly sampled from Gaussian distributions defined by their respective uncertainties. A least-squares fit was then applied to the logarithm of the molecule’s desorption rate versus the inverse of the temperature, optimizing all five experimental curves simultaneously to derive the binding energy ($E_b$). This transformation, known as an Arrhenius plot, allows the data to be fitted with a straight line, thus mitigating temperature-dependent fitting biases that can arise when applying Monte Carlo sampling to exponential trends.

This analysis yields a mean best-fit \ce{H2S}$-$\ce{H2S} binding energy of E$_b$ = $3159\pm46$ K, shown in the original exponential format by the blue curves in Figure \ref{fig:TPD_multi_leadingedge}. The log-transformed Arrhenius fits are shown in Appendix \ref{Appendix:multiL_arrhenius}. The uncertainties in E$_b$ are primarily driven by errors in the absolute substrate temperature and in $\nu_{\text{TST}}$, as the errors in ice coverage and from the fitting are $\lesssim$1 K. Given that the uncertainties in $\nu_{\text{TST}}$ due to temperature readout errors are small ($\sim$1\%), we adopt the standard deviation from the five experiments (7$\times$10$^{14}$ ML s$^{-1}$, or $\sim$4\%) as the $\nu_{\text{TST}}$ error in the fits, which yields uncertainties of $\lesssim$5 K in the binding energy. Consequently, most of the uncertainty in E$_b$ are due to the absolute temperature error. Another approach could be to adopt as uncertainties in $\nu_{\text{TST}}$ the full range of values derived for the temperatures within the limits encompassed by the fit. However, this also does not affect the results appreciably: while, in this case, $\nu_{\text{TST}}$ uncertainties correspond to approximately 20\%, the resulting mean E$_b$ value remains unchanged, and the error is only increased by 1 K. Additionally, it is also possible to incorporate a temperature-dependent $\nu_{\text{TST}}$ value into the fit, but this approach produces a poorer fit to the data and was thus not preferred (see Appendix \ref{Appendix:T_dependent_nuTST}). In some instances, small artifacts are observed at the leading edge, and occasionally symmetrically at the trailing edge of the TPD curves. However, these artifacts have an insignificant impact on the results, with masked fits yielding E$_b$ values that differ by less than 1 K. Table \ref{tab:rec_E_nu} summarizes the recommended desorption parameters derived experimentally for the multilayer \ce{H2S} ice. The recommended preexponential factor is the mean value between all five thicknesses, with its uncertainty corresponding to their standard deviation: $\nu_\text{TST}$ = ($1.76\pm0.07$)$\times$10$^{16}$ ML s$^{-1}$.

\begin{table*}[htb!]
\centering
\caption{Summary of the \ce{H2S} desorption parameters derived experimentally in this work}
\label{tab:rec_E_nu} 
\begin{tabular}{lccccc}  
\toprule\midrule
Regime                  &   Substrate       &   n   &   Coverage    &   E$_b$                           &   $\nu_\text{TST}$\\
                        &                   &       &   (ML)        &   (K)                             &   (ML$^{1-n}$ s$^{-1}$)\\      
\midrule
\ce{H2S}$-$\ce{H2S}     &   \ce{CsI} window &   0   &   19$-$90     &   $3159\pm46$                     &   ($1.76\pm0.07$)$\times$10$^{16}$\\
\midrule
\ce{H2S}$-$\ce{H2O}     &   cASW            &   1   &   0.6         &   3355 [179] $^a$                 &   ($1.69\pm 0.04$)$\times$10$^{16}$\\
\ce{H2S}$-$\ce{H2O}     &   cASW            &   1   &   0.5         &   3368 [172] $^a$                 &   ($1.72\pm 0.04$)$\times$10$^{16}$\\
\ce{H2S}$-$\ce{H2O}     &   cASW            &   1   &   0.3         &   3401 [169] $^a$                 &   ($1.83\pm 0.05$)$\times$10$^{16}$\\
\ce{H2S}$-$\ce{H2O}     &   cASW            &   1   &   0.2         &   3443 [191] $^a$                 &   ($1.85\pm 0.05$)$\times$10$^{16}$\\
\bottomrule
\multicolumn{6}{l}{\footnotesize{$^a$ Square brackets denote the FWHMs of the binding energy distributions.}}\\
\end{tabular}
\end{table*}

\subsubsection{\ce{H2S}$-$\ce{H2O}}
\label{sec:BE_subML}

\begin{figure}[htb!]\centering
\includegraphics[scale=0.27]{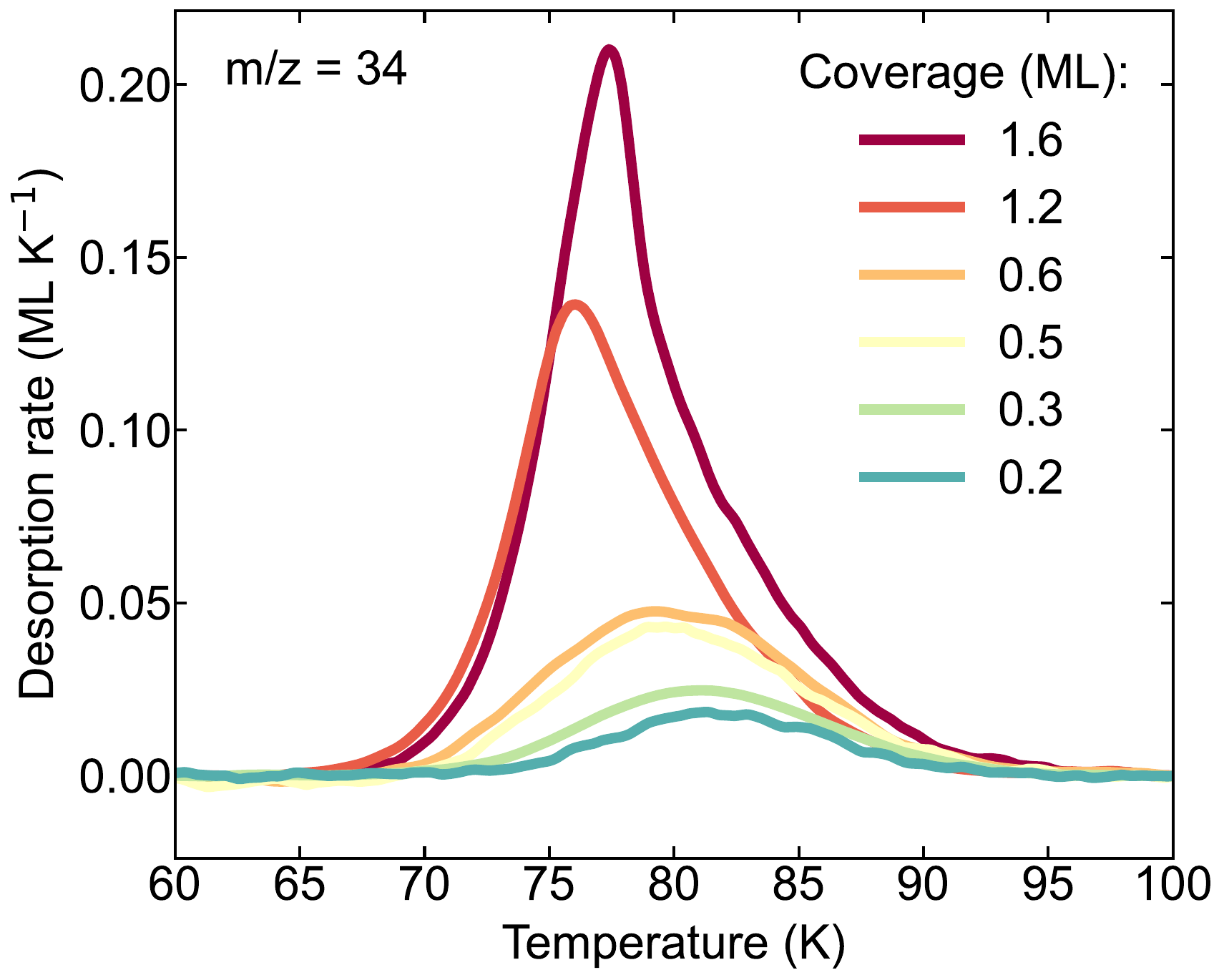}
\caption{TPD curves of the low-coverage \ce{H2S} ice experiments deposited on top of a cASW substrate. The transition from zeroth-order to pure first-order sublimation behavior is seen for coverages $<$1.2 ML. The experimental data is smoothed for clarity.}
\label{fig:subML_TPDs}
\end{figure}

Figure \ref{fig:subML_TPDs} shows the TPD curves of the low-coverage \ce{H2S} ice grown on top of a cASW substrate. In total, six coverages are depicted: 1.6, 1.2, 0.6, 0.5, 0.3, and 0.2 ML (see Appendix \ref{Appendix:subML_calib} for details on the submonolayer coverage estimation). The two largest thicknesses, 1.6 and 1.2 ML, show nearly overlapping leading edges that culminate in sharp peaks at $\sim$77 and $\sim$76 K, respectively---in accordance with the expected dominating zeroth-order desorption behavior. In contrast, the submonolayer coverages display a markedly different profile, with misaligned leading edges and peak desorption temperatures that increase slightly with decreasing coverages (T$_\text{peak}$ $\sim$ 79.3, 79.6, 81.1, and 81.3 K for 0.6, 0.5, 0.3, and 0.2 ML). This behavior is consistent with pure first-order desorption, and therefore the 0.6$-$0.2 ML coverages are used to derive the \ce{H2S}$-$\ce{H2O} binding energy (dominated by \ce{H2S} interactions with water).

\begin{figure}[htb!]\centering
\includegraphics[scale=0.28]{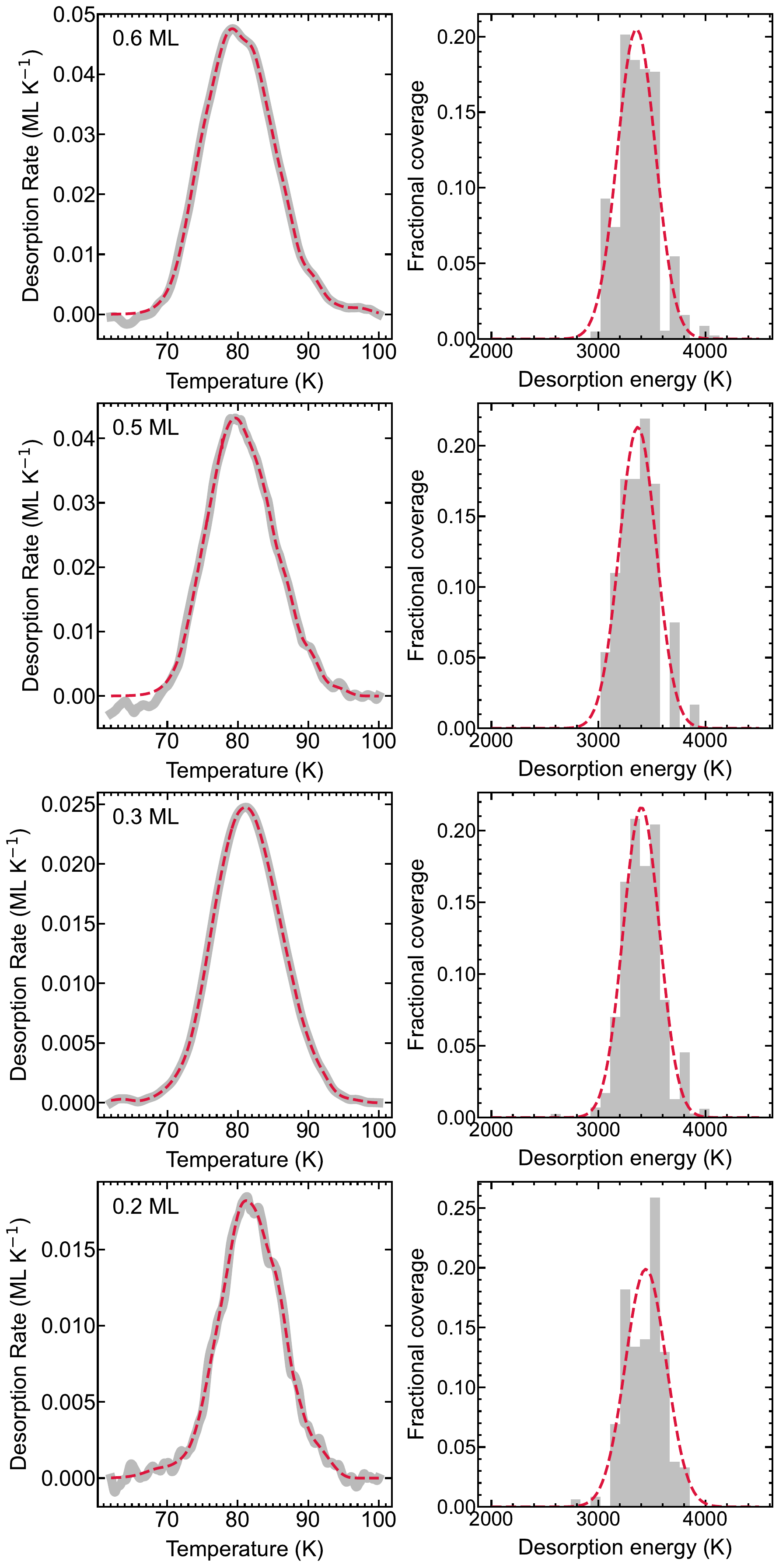}
\caption{Results of the fits to the submonolayer \ce{H2S} ice data. Each row corresponds to a different \ce{H2S} ice thickness: $\sim$0.6, $\sim$0.5, $\sim$0.3, and $\sim$0.2 ML (from top to bottom). Left panels show the TPD curves with the experimental data in gray and the fits with a linear combination of first-order Polanyi-Wigner equations in red. Right panels show the corresponding binding energy distribution: normalized fractional coverages are shown as a functional of the binding energy (gray histogram), superimposed by a Gaussian fit to the distribution (red dashed line). }
\label{fig:fits_subML_all}
\end{figure}

In the case of the submonolayer TPD curves, unlike the multilayer curves, the entire desorption profile is fitted to find E$_b$. As a result, the peak desorption temperature of the curve is encompassed by the model, and thus we calculate $\nu_{\text{TST}}$ at the peak value for each coverage---which reflects the transition state of the largest parcel of desorbed \ce{H2S} molecules. Also differently from the multilayer regime, the TPD curves of the submonolayer \ce{H2S} ice cannot be described by a single binding energy (see Appendix \ref{Appendix:subML_oneE} for an example of an attempted fit). This is a direct consequence of the non-homogeneous nature of the cASW substrate, which generates multiple binding sites that result in a range of \ce{H2S}$-$\ce{H2O} desorption energies. To account for that, we fit a linear combination of first-order Polanyi-Wigner equations (eq. \ref{eq:Polanyi}) to the curves, with statistical weights normalized to the initial ice coverages for each experiment. This allows us to determine the relative population of each binding site. The same approach has been used in the past to derive submonolayer binding energies of hypervolatiles such as \ce{N2} and \ce{CO}, as well as less volatile molecules such as 2-C and 3-C hydrocarbons and methanol, on varying substrates (from water ice to graphite; \citealt{Doronin2015, Fayolle2016, Behmard2019}). Since the preexponential factor is calculated using equations \ref{eq:nuTST}, \ref{eq:qTR}, and \ref{eq:qROT}, it is treated as a fixed parameter when fitting the curves with a linear combination of Polanyi-Wigner equations. Our sampled desorption energies range from 2000 to 4500 K in steps of 90 K, chosen to balance fit resolution and degeneracies associated with smaller bin sizes. Figure \ref{fig:fits_subML_all} shows the resulting fits to the submonolayer TPD curves and their corresponding binding energy distributions, plotted as fractional coverages as a function of the energy. 

The population histograms can be approximated by a normal distribution and are thus fitted by a Gaussian curve to allow for more straightforward interpretation (red dashed line in the right panels in Figure \ref{fig:fits_subML_all}). The mean binding energy and FWHM values for each coverage are listed in Table \ref{tab:rec_E_nu} and can be regarded as representative desorption energies for each ice thickness. The corresponding $\nu_\text{TST}$ values used in the fits are also listed, with their uncertainties stemming from the absolute error in the substrate temperature. Among the four coverages, the average binding energy and preexponential factor are $E_b=3392\pm 56$ K and $\nu_\text{TST}=(1.7\pm 0.1)\times10^{16}$ s$^{-1}$. The uncertainty of $E_b$ encompasses the relative error due to the step size in the binding energy sampling and the standard deviation among the four coverages, while for $\nu_\text{TST}$ it is taken solely from its standard deviation. 

\subsubsection{Binding energies vs coverage}

\begin{figure}[htb!]\centering
\includegraphics[scale=0.43]{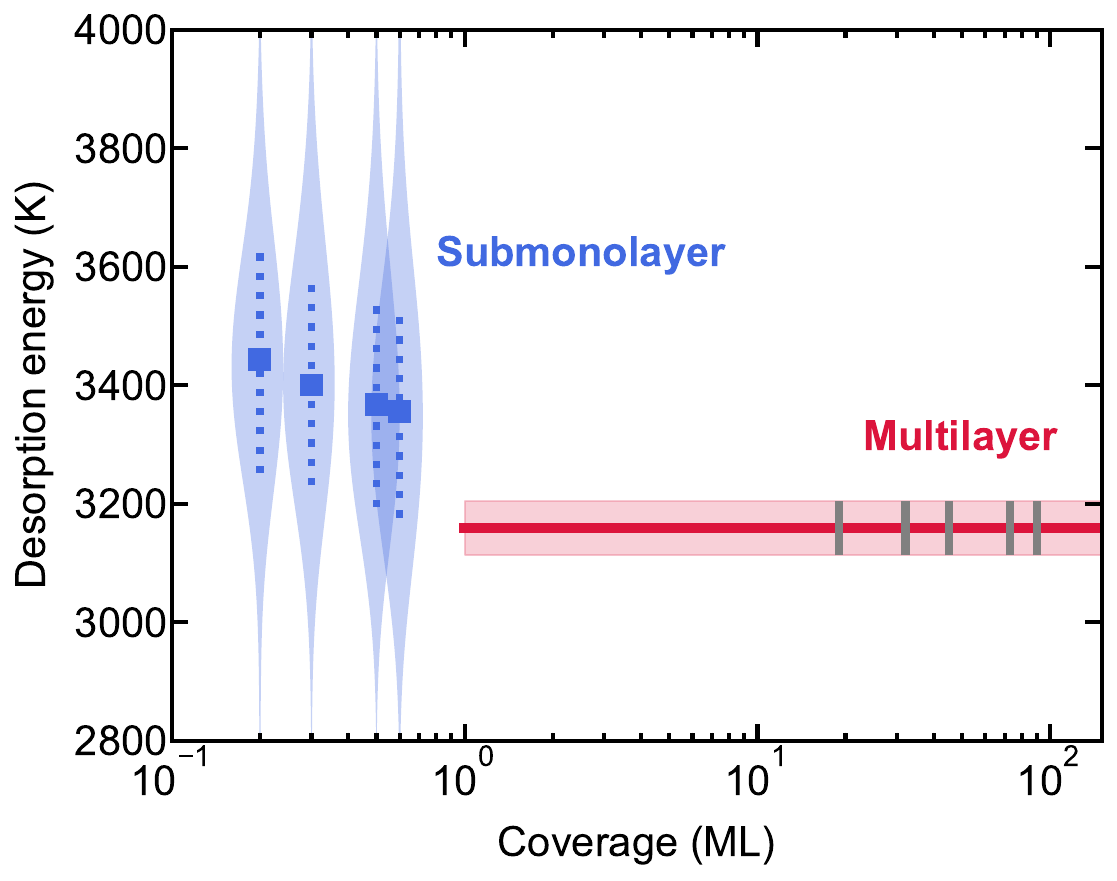}
\caption{Desorption energies derived experimentally in this work as a function of \ce{H2S} ice thickness. For the submonolayer regime, the binding energies are represented by blue violin plots, where the square markers indicates the mean binding energies, and the vertical dotted lines show the FWHMs of the Gaussian fits to the binding energy distributions for each coverage. The contour of each violin plot reflects the full range of the sampled binding energy distribution, with its thickness normalized according to the statistical weights (or fractional coverages) of each E$_b$ value. The multilayer binding energy is depicted as a red solid line with its uncertainty shown by the light-red shadowed area. The five different ice coverages used to derive the multilayer E$_b$ are shown as gray vertical lines.}
\label{fig:Edes_coverage}
\end{figure}

In Figure \ref{fig:Edes_coverage}, we show a comparison of all \ce{H2S} binding energies derived in this work as a function of coverage. The multilayer binding energy ($E_b\sim3159$ K) is slightly lower than the submonolayer values, with a difference of $\sim$7\% w.r.t the average submonolayer E$_b$ ($\sim3392$ K). This small difference suggests that the \ce{H2S}$-$\ce{H2S} interactions are only moderately weaker than the \ce{H2S}$-$\ce{H2O} counterparts. This is rather expected: both \ce{H2S} and \ce{H2O} interact via hydrogen bonding networks---known as one of the strongest intermolecular forces---where hydrogen atoms are covalently bonded to an electronegative atom. Since sulfur is larger than oxygen, it is less electronegative, and therefore the H-bonding networks are weaker for \ce{H2S} than for \ce{H2O}. Indeed, the fact that pure first-order desorption could be achieved at coverages of $\sim$0.6 ML (as evinced by the profile of the TPD curves in Figure \ref{fig:subML_TPDs}, see section \ref{sec:BE_subML}) suggests that wetting of the \ce{H2S} on the cASW surface proceeds relatively uniformly, as opposed to having a tendency to form \ce{H2S} islands (for comparison, \citealt{Bergner2022} found that a dose of 0.04 ML was required to achieve the first-order desorption regime for \ce{HCN} on cASW). This is in line with a (modest) preference of \ce{H2S} to interact with \ce{H2O}. 

In the submonolayer regime, the mean \ce{H2S}$-$\ce{H2O} binding energy increases with decreasing ice coverage, in line with other laboratory measurements for various adsorbate-substrate combinations (e.g., \citealt{Noble2012, Fayolle2016, He2016b, Nguyen2018, Behmard2019}). This generalized phenomenon can be explained by two effects: 1. lateral interactions influencing the binding energy of the adsorbates, or 2. a preference of the adsorbates to occupy deeper binding sites (caused by their diffusion from shallower sites)---with the most likely explanation being a combination of the two. The weaker \ce{H2S}$-$\ce{H2S} interactions relative to \ce{H2S}$-$\ce{H2O} may manifest as a slight decrease in binding energy caused by lateral \ce{H2S} interactions compared to isolated \ce{H2S} adsorbates fully interacting with \ce{H2O}, in support of explanation 1. Complementarily, the alignment of the trailing edges in the submonolayer experiments suggests that the deeper binding sites are similarly occupied for all explored thicknesses, reinforcing explanation 2.

To the best of our knowledge, this is the first study to experimentally determine binding energies for \ce{H2S} ice analogues. Nonetheless, estimations based on laboratory data have been proposed in the past. \cite{Minissale2022} recommend a value of E$_b$ = 3426 K for \ce{H2S} on a cASW substrate based on the peak desorption temperature of \ce{H2S} and the associated $\nu$ obtained using the TST formalism---in good agreement with our measurements. In contrast, the E$_b$ value of 2296$\pm$9 K estimated by \cite{Penteado2017}, based on the relative peak desorption temperature of \ce{H2S} with respect to \ce{H2O} and the binding energy of the latter, differs significantly from our measurements. Computational efforts have also been made to estimate the binding energies of \ce{H2S}. In general, computed E$_b$ distributions for \ce{H2S} on water substrates range between $\sim2000-3600$ K \citep{Wakelam2017, Das2018, Ferrero2020, Piacentino2022}, but some studies predict much lower ranges, with upper limits closer to 2000 K \citep{Oba2018, Bariosco2024}. The experimental values reported here for \ce{H2S}$-$\ce{H2O} (and \ce{H2S}$-$\ce{H2S}) are generally not well reproduced by the computations, often falling within their upper bounds or, in some cases, being entirely underpredicted. This discrepancy is likely due to the limitations in the computational methods to incorporate diffusion as part of their binding energy calculations, which can lead to systematic underestimations by failing to account for the tendency of adsorbates to settle into deeper binding sites prior to sublimating. In an astrophysical context, icy mantles shrouding dust grains are gradually heated by protostellar radiation, in which case adsorbates in shallow sites become free to diffuse throughout the ice and find deeper binding sites before eventually sublimating. The binding energies derived from TPD experiments are therefore expected to better reproduce the conditions in space.

\subsection{Entrapment in \ce{H2O}}

\begin{figure}[htb!]\centering
\includegraphics[scale=0.25]{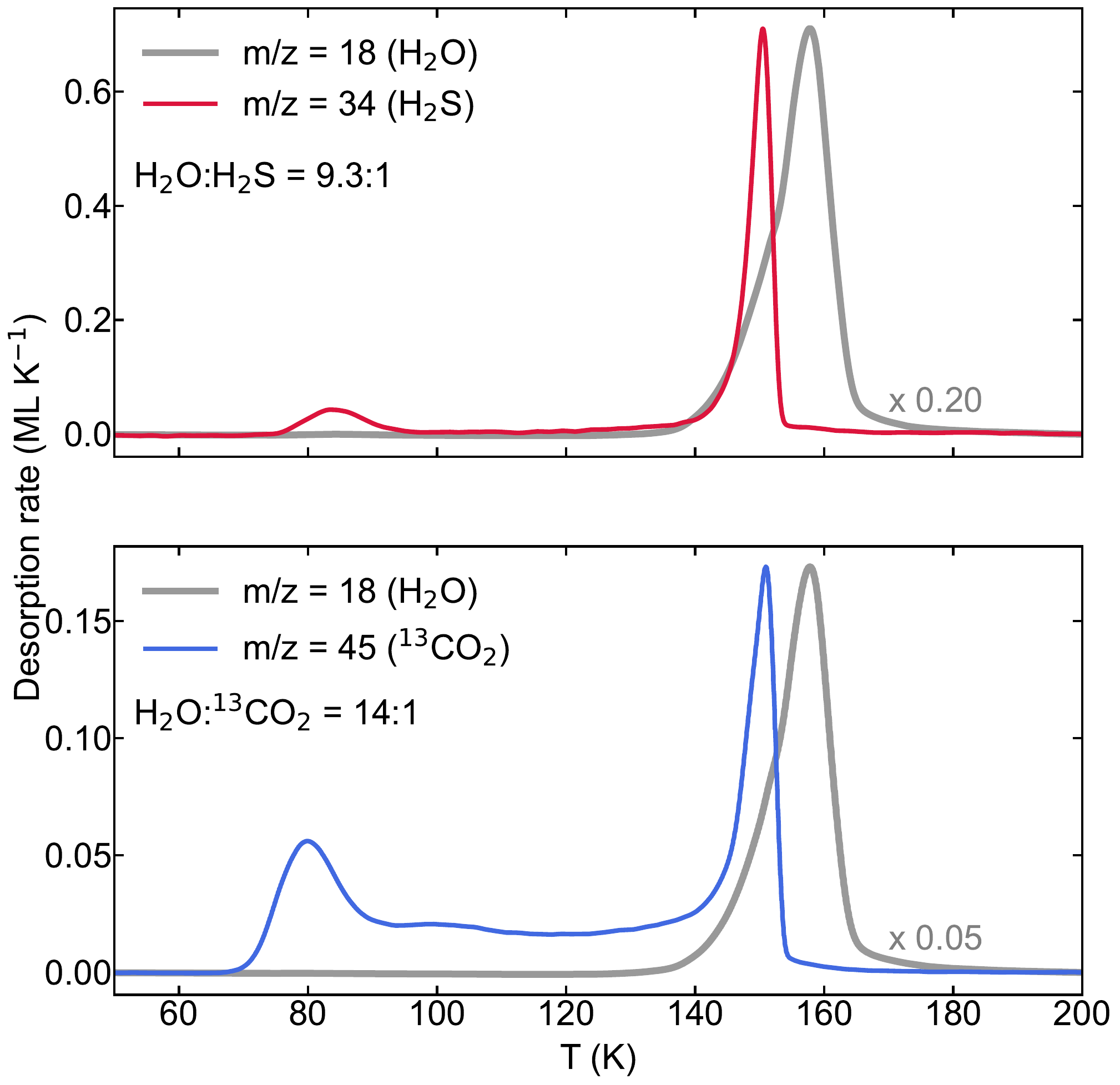}
\caption{TPD curves of two mixed-ice experiments, \ce{H2O}:\ce{H2S} = 9.3:1 (top) and \ce{H2O}:\ce{^{13}CO2} = 14:1 (bottom). The desorption data of \ce{H2O}, \ce{H2S}, and \ce{^{13}CO2} are shown in gray (m/z = 18, \ce{[H2O]^+}), red (m/z = 34, \ce{[H2S]^+}), and blue (m/z = 45, \ce{[^{13}CO2]^+}), respectively. Two desorption regimes are seen for \ce{H2S}: a submonolayer \ce{H2S}$-$\ce{H2O} desorption feature and a molecular volcano desorption peak. For \ce{^{13}CO2}, a third regime is observed between its submonolayer and volcano peaks, indicating constant sublimation within that temperature range.}
\label{fig:entrapment_TPDs}
\end{figure}

The entrapment behavior of \ce{H2S} in water-dominated ice was investigated by TPD experiments of \ce{H2O}:\ce{H2S} ice mixtures with four different ratios (5.1:1, 7.5:1, 9.3:1, and 17:1, determined from the final IR spectra measured after deposition) and constant total ice thicknesses ($\sim$40 ML). These mixing conditions are chosen to ensure that \ce{H2S} is mostly surrounded by \ce{H2O} molecules while still allowing for measurements with good signal-to-noise ratios, particularly for the more water-dominated cases. Additionally, a control experiment consisting of \ce{^{13}CO2} mixed in \ce{H2O} ice with a ratio of \ce{H2O}:\ce{^{13}CO2}=14:1 was also performed with the goal of testing the potential role of intermolecular interactions on the entrapment efficiencies. Both \ce{H2S} and \ce{CO2} have similar volatilities, with desorption temperatures typically falling in similar ranges for the same pressure and substrate conditions. However, \ce{CO2} is an apolar molecule and its intermolecular interactions with \ce{H2O} are markedly different from those of \ce{H2S}. The resulting TPD curves of the \ce{H2O}:\ce{H2S} = 9.3:1 experiment, as well as the control \ce{H2O}:\ce{^{13}CO2} = 14:1 experiment, are shown in the upper and lower panels of Figure \ref{fig:entrapment_TPDs}, respectively.

Two desorption features are observed for \ce{H2S} ice mixed in \ce{H2O}: a monolayer desorption peak at $\sim$84 K, and a molecular volcano feature at $\sim$150 K. The former is consistent with \ce{H2S} desorption characterized by the \ce{H2S}$-$\ce{H2O} binding energy (see Appendix \ref{Appendix:TPD_mixed_ices}), and includes contributions from \ce{H2S} adsorbates on the uppermost ice layer, as well as \ce{H2S} molecules occupying channels within the \ce{H2O} matrix with access to the surface. The molecular volcano is the strongest desorption peak, and corresponds to \ce{H2S} being released from the \ce{H2O} ice matrix as the ice crystallizes and cracks, creating new channels to the surface. In the case of \ce{^{13}CO2}, a third desorption regime is observed ranging from $\sim90-140$ K. Following its monolayer desorption feature at $\sim$80 K, the TPD curve does not return to zero desorption rate, but instead shows a relatively steady desorption of \ce{^{13}CO2} until the molecular volcano feature at $\sim$150 K---after which it appears to be completely sublimated. This continuous desorption between the two main features might be related to the diffusion of \ce{^{13}CO2} through the \ce{H2O} ice. 

\begin{figure}[htb!]\centering
\includegraphics[scale=0.4]{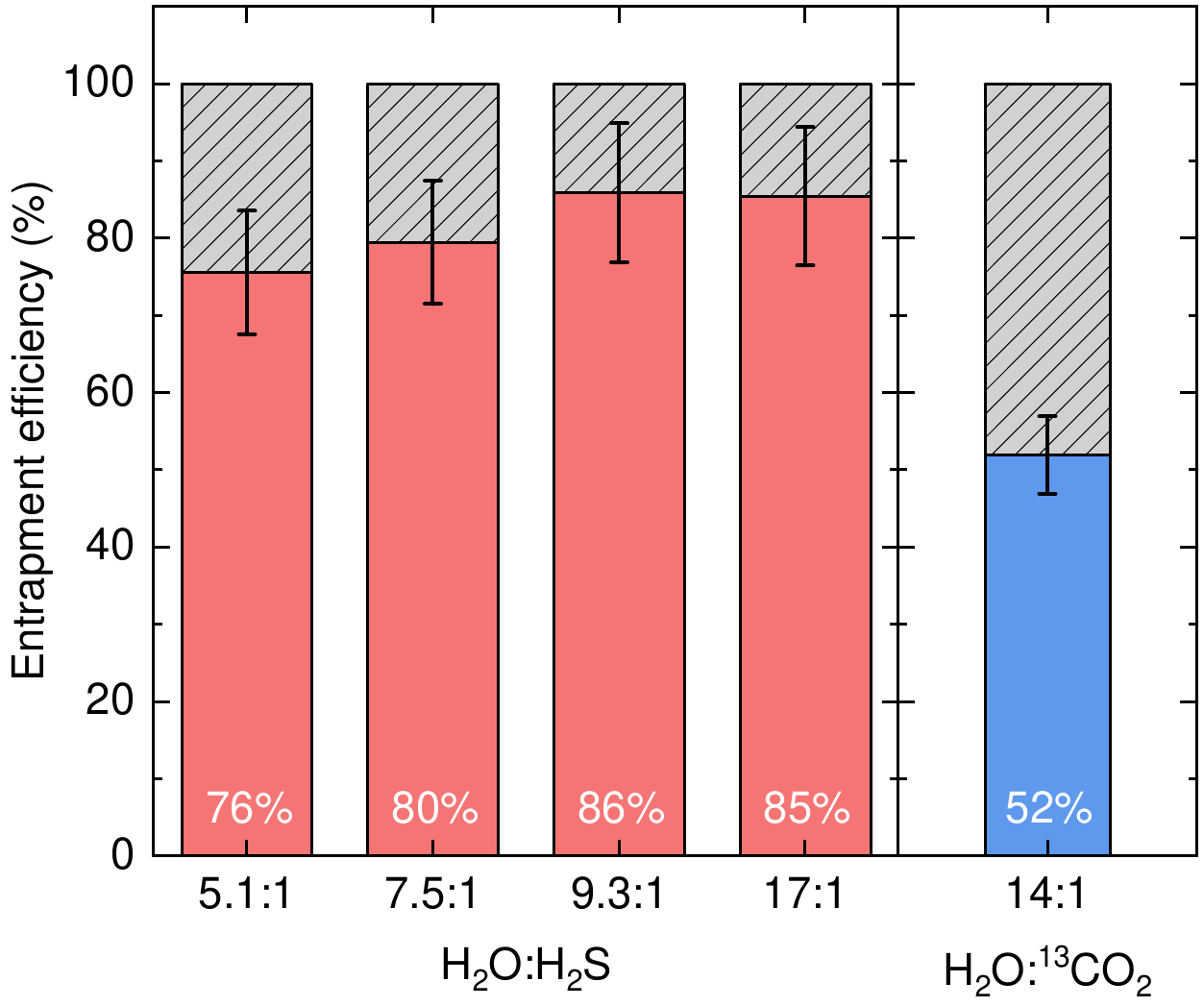}
\caption{Efficiencies with which \ce{H2S} (red) and \ce{^{13}CO2} (blue) are trapped in water-rich binary ice mixtures as derived from our experiments (see also Table \ref{tab:trapping}).}
\label{fig:all_trapping}
\end{figure}

Table \ref{tab:trapping} lists and Figure \ref{fig:all_trapping} depicts the entrapment efficiencies of \ce{H2S} and \ce{^{13}CO2} measured from our mixed-ice experiments (see section \ref{sec:2_analysis} for details on how these are determined). The uncertainties are dominated by experimental variability and are thus taken as 10\% of the entrapment efficiency, based on the estimations derived by \cite{Simon2023} with the same experimental setup for \ce{H2O}:hypervolatile mixtures with ratios 3:1. In comparison, the error due to stochastic instrumental noise amounts to $\lesssim$2\%. Overall, \ce{H2S} experiences very efficient entrapment in the \ce{H2O}-ice matrix, with $\sim76-85\%$ of the \ce{H2S} unable to escape to the gas phase until \ce{H2O} undergoes significant structural changes. This efficient entrapment of \ce{H2S} has been alluded to in the past by \cite{Jimenez-Escobar2011} based on experiments using \ce{H2O}:\ce{H2S} ice mixtures with roughly 10\% \ce{H2S} concentrations. Larger water abundances relative to \ce{H2S} result in higher entrapment efficiencies, consistent with the behavior observed for many other species mixed in \ce{H2O} and \ce{CO2}-dominated ices \citep{Fayolle2011, Martin-Domenech2014,Simon2019, Simon2023, Kruczkiewicz2024, Pesciotta2024}. \cite{Fayolle2011} propose two explanations for this phenomenon: 1. there is a reduction in porous channels connected to the surface for higher dilutions of the volatile; and/or 2. higher volatile concentrations increase its diffusion lengths. The two highest dilutions (\ce{H2O}:\ce{H2S} $\sim$9:1 and 17:1) yield analogous efficiencies within their uncertainties, signaling that the trapping capacity of \ce{H2S} in the \ce{H2O} ice saturates at $\sim$85\% for our deposition conditions and a total ice thickness of $\sim$40 ML. Similarly, \cite{Pesciotta2024} recently reported a leveling-off of the entrapment efficiencies of \ce{CO} in both \ce{H2O} and \ce{CO2} dominated binary ice mixtures starting at concentrations of $\lesssim$1:10 volatile:matrix. We emphasize, however, that the \ce{CO} entrapment saturation measured in their work occurred at significantly lower efficiencies than the saturation point for \ce{H2S} in \ce{H2O} measured here (they find $\sim$69\% and $\sim$61\% for \ce{CO2} and \ce{H2O} matrices, respectively, in 1:15 \ce{CO}:matrix ratios and ice thicknesses $\gtrsim$50 ML).


\begin{table}[htb!]
\centering
\caption{Summary of the entrapment efficiencies derived experimentally in this work. All experiments kept a constant total ice thickness of $\sim$40 ML.}
\label{tab:trapping} 
\begin{tabular}{lccc}  
\toprule\midrule
Mixture                 &   Ratio   &   Entrapment efficiency\\
                        &           &   (\%)\\
\midrule
\ce{H2O}:\ce{H2S}       &   5.1:1   &   76$\pm$8\\
\ce{H2O}:\ce{H2S}       &   7.5:1   &   80$\pm$8\\ 
\ce{H2O}:\ce{H2S}       &   9.3:1   &   86$\pm$9\\ 
\ce{H2O}:\ce{H2S}       &   17:1    &   85$\pm$9\\ 
\ce{H2O}:\ce{^{13}CO2}  &   14:1    &   52$\pm$5\\ 
\bottomrule
\end{tabular}
\end{table}

In comparison to \ce{H2S}, \ce{^{13}CO2} is significantly less entrapped in the \ce{H2O} ice, with an efficiency of $\sim52\%$ for a \ce{H2O}:\ce{^{13}CO2} ratio of $\sim$14:1. This stark difference between two similarly volatile molecules might be explained by the nature of their interactions with the water matrix. For \ce{^{13}CO2}, the oxygen atoms may serve as hydrogen bond acceptors; however, their very low partial negative charges result in exceptionally weak hydrogen bonds \citep{Zukowski2017}. Interactions between \ce{^{13}CO2} and \ce{H2O} are thus primarily governed by weaker van der Waals forces. In contrast, the hydrogen bonds between \ce{H2S} and \ce{H2O}, while not as strong as \ce{H2O}$-$\ce{H2O} counterparts \citep{Craw1992}, still offer significant energetic stabilization. This is reflected in the mean \ce{H2S}$-$\ce{H2O} binding energy measured here at 3392$\pm$56 K, which surpasses the \ce{CO2}$-$\ce{H2O} binding energies measured in the literature by $\sim$50$-$60\%\footnote{\cite{Noble2012} and \cite{Edridge2013} measured the mean monolayer \ce{CO2} binding energy on ASW to be $\sim$2247 K and $\sim$2100 K, respectively. The former was measured on a nonporous ASW substrate, while the latter was measured on porous ASW.}. In fact, both \cite{Noble2012} and \cite{Edridge2013} reported \ce{CO2}$-$\ce{CO2} binding energies to be \textit{higher} than the mean value for those of \ce{CO2}$-$\ce{H2O}, in clear contrast with the behavior observed here for \ce{H2S} and water. The difference in the interaction energetics between the two systems (\ce{H2O}:\ce{^{13}CO2} vs \ce{H2O}:\ce{H2S}) may result in \ce{H2S} diffusing less readily through \ce{H2O} ice compared to \ce{^{13}CO2}, leading to more efficient trapping of \ce{H2S}. Indeed, the steady desorption of \ce{^{13}CO2} between its monolayer and volcano peaks may suggest that it is more mobile than \ce{H2S}, for which this behavior is minimal. Specifically, this temperature range accounts for $\lesssim5\%$ of the integrated QMS signal in all \ce{H2S} experiments, and corresponds to only $\sim2\%$ in the \ce{H2O}:\ce{H2S} = 9.3:1 mixture, compared to $\sim21\%$ for \ce{^{13}CO2} in a higher dilution.  

In addition to intermolecular interactions, the mass and size of the molecule may also influence its diffusion and entrapment. Based on their molecular masses and kinetic diameters, \ce{^{13}CO2} is $\sim$30\% heavier and $\sim$9\% smaller than \ce{H2S} ($m_{\ce{^{13}CO2}}\sim45$ amu vs $m_{\ce{H2S}}\sim34$ amu; $d_{\ce{^{12}CO2}}\sim3.3\AA$ vs $d_{\ce{H2S}}\sim3.6\AA$, \citealt{Matteucci2006, Ismail2015}). Considering the mass effect alone, the attempt frequency for diffusion of \ce{^{13}CO2} would be smaller than that of \ce{H2S}, resulting in the former diffusing slower (see, e.g., \citealt{Cuppen2017}) and consequently being more efficiently trapped. The fact that the inverse is observed in our experiments signals that the mass effect is overpowered by other factors favoring the entrapment of \ce{H2S}. In contrast, the size difference could result in \ce{H2S} molecules being more efficiently trapped if the mean pore size is smaller than the kinetic diameter of \ce{H2S}, but larger than that of \ce{\ce{^{13}CO2}}. This size mismatch may create a geometrical limitation, with the pores acting similarly to a net that restricts diffusion. For instance, gas permeance through microporous silica membranes (pore size between $3.8-5.5 \AA$) has been shown to decrease steeply with the species' kinetic diameter, with the permeance of \ce{N2} ($d\sim3.6 \AA$, similar to \ce{H2S}; \citealt{Ismail2015}) being roughly one order of magnitude lower than that of \ce{CO2}  \citep{DeVos279}. ASW ices are typically found to be microporous (pore width $\leq20 \AA$), with no lower limits estimation for the pore sizes and little to no incidence of mesopores \citep{Mayer1986, Raut2007, Cazaux2015, Carmack2023}. This is particularly true for ices grown through collimated deposition---the technique employed in this work. The possibility of a pore effect therefore cannot be ruled out. Pore effects could also play a role in the constant sublimation regime of \ce{CO2}, observed between its monolayer and volcano features. The same behavior is seen for \ce{^{12}CO2}:\ce{H2O} ice mixtures in \cite{Kruczkiewicz2024}, though they note that this is not observed for other, more volatile species (such as Ar and CO) trapped in \ce{H2O}-dominated binary ices. Since both Ar and \ce{CO} have larger kinetic diameters than \ce{CO2} ($d_{\ce{Ar}}\sim3.4\AA$ and $d_{\ce{CO}}\sim3.8\AA$; \citealt{Breck1973, Matteucci2006}), the fact that they do not display the constant sublimation regime in the experiments by \cite{Kruczkiewicz2024} does not preclude this possibility. Dedicated experimental investigations are necessary to fully constrain the effect of the species' size and intermolecular interactions to entrapment. Nonetheless, the overall result is that \ce{H2S} is very efficiently entrapped in \ce{H2O} ice, more so than other similarly volatile molecules, which has significant implications to its gas vs ice distribution in planet-forming regions.

\section{Astrophysical implications}\label{sec:astro}

The binding energies derived experimentally in this study are used to estimate the locations of the \ce{H2S} sublimation fronts in the midplane of a representative protoplanetary disk. First, we calculate the desorption temperatures of \ce{H2S} for its two binding energies (\ce{H2S}$-$\ce{H2S} and \ce{H2S}$-$\ce{H2O}) following the formalism by \cite{Hollenbach2009}, in which the desorption temperature is found by equating the flux of molecules adsorbing on the grain surface to the flux of molecules desorbing from it:

\begin{equation}
    T_i\sim(E_{b,i}/k)\left[ \ln{\left( \frac{4N_if_i\nu}{n_iv_i}\right)}\right]^{-1},
\end{equation}

\noindent where $T_i$ is the sublimation temperature of species $i$, $E_{b,i}$ is its binding energy, $N_i$ is the number of adsorption sites (assumed to be $\sim$10$^{15}$ cm$^{-2}$), $f_i$ is the fraction of such sites occupied by species $i$, and $\nu_i$ is its vibrational frequency in the surface potential well (i.e., the preexponential factor in equation \ref{eq:Polanyi}). In the denominator, $n_i$ is the number density of species $i$ in the gas phase, and $v_i$ is its thermal speed. We estimate $f_i$ based on the cometary abundance of \ce{H2S} relative to water, \ce{H2S}/\ce{H2O} = 1.10$\pm$0.05\%, as measured by the Rosetta mission on the coma of comet 67P \citep{Calmonte2016}; and assuming a cometary composition of $\sim80\%$ water (e.g., \citealt{Deslsemme1991}). The number density $n_i$ is estimated by multiplying the cometary \ce{H2S}/\ce{H2O} by the water abundance with respect to \ce{H} and the number density of hydrogen nuclei in a disk midplane (taken as $\ce{H2O}/\ce{H}\sim10^{-4}$, e.g., \citealt{Drozdovskaya2015, Boogert2015}; and $n_H\sim10^{10}$ cm $^{-3}$, e.g., \citealt{Walsh2014}). This yields sublimation temperatures for \ce{H2S}$-$\ce{H2S} and \ce{H2S}$-$\ce{H2O} of $\sim$64 K and $\sim$69 K, respectively. For the latter, the mean \ce{H2S}$-$\ce{H2O} binding energy value was used (E$_b \sim 3392$ K). We note that the desorption temperatures in the disk model differ from those in the laboratory due to variations in physical conditions between the two environments.

\begin{figure}[htb!]\centering
\includegraphics[scale=0.42]{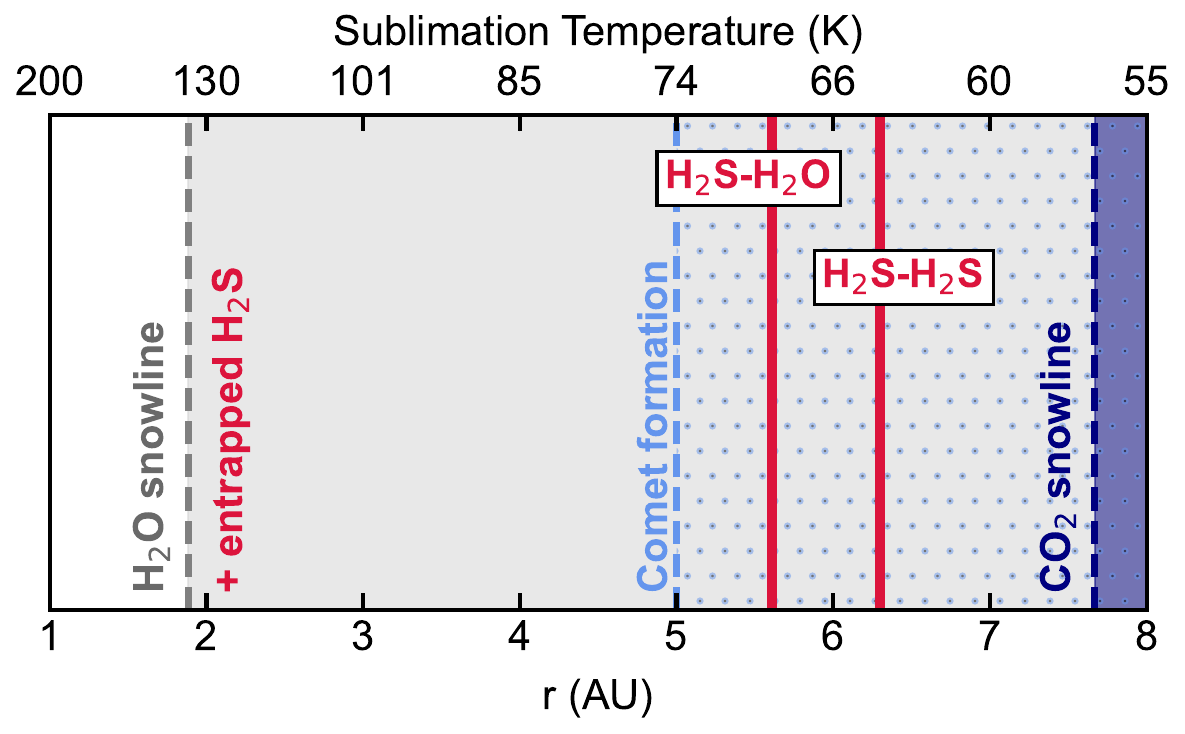}
\caption{Sublimation front locations for the \ce{H2S}$-$\ce{H2S} and \ce{H2S}$-$\ce{H2O} binding energies estimated for a representative T-Tauri disk midplane. Most of the \ce{H2S} will remain entrapped in the ice past its sublimation front radii, and will only be released to the gas phase close to the water snowline. The \ce{CO2} snowline location is also shown for comparison. It was estimated based on the mean binding energy derived by \cite{Noble2012} for (sub)monolayer \ce{CO2} ice on a nonporous ASW surface and their adopted preexponential factor of 10$^{12}$ s$^{-1}$, while assuming ice abundances $N$(\ce{CO2})/$N$(\ce{H2O})$\sim$0.28 (median value observed in low-mass young stellar objects; \citealt{Boogert2015}).}
\label{fig:subl_fronts}
\end{figure}

To derive the locations of the sublimation fronts, we assume a representative disk midplane radial temperature profile:
\begin{equation}
    T=200\times r^{-0.62},
\end{equation}
\noindent where $r$ is the radial distance in a.u. This corresponds to the median midplane temperature distribution found for a sample of 24 T-Tauri disks \citep{Andrews2007}. The resulting sublimation front locations are shown in Figure \ref{fig:subl_fronts}.

The sublimation fronts for the \ce{H2S}$-$\ce{H2S} and \ce{H2S}$-$\ce{H2O} binding energies occur at $\sim$6.3 au and $\sim$5.6 au, respectively. The former corresponds to pure \ce{H2S} ice desorption, while the latter is relevant for \ce{H2S} desorbing from a \ce{H2O}-rich surface. In the case of \ce{H2S} entrapped in water ice, its sublimation will be delayed to much closer to the water snowline at $\sim$1.9 au (assuming $E_{b,\ce{H2O}}\sim5600$ K, \citealt{Wakelam2017}; and a typical $\nu\sim10^{13}$ s$^{-1}$). The precise location of this sublimation front will depend on the water crystallization kinetics, but will generally occur at slightly lower temperatures than \ce{H2O} desorption, placing it just beyond the \ce{H2O} snowline.

The deuterium fractionation of gaseous \ce{H2S} observed towards Class 0 protostars suggests that it is formed in ices during early prestellar cloud timescales, before the onset of the catastrophic CO freeze-out \citep{Ceccarelli2014}. Most of the water ice is also formed during similarly early timescales, resulting in \ce{H2S} likely inhabiting a \ce{H2O}-rich ice environment---the so-called polar ice layer. The sublimation behavior of \ce{H2S} will therefore be likely dominated by its entrapment within \ce{H2O} ice, given its high efficiency as measured in this work ($\sim$85\% for \ce{H2S} concentrations of $\sim$5$-$10\% in \ce{H2O}). In more representative interstellar scenarios, the bulk \ce{H2S} concentration in \ce{H2O} ice is expected to be considerably smaller ($\lesssim$1\% based on observationally-constrained \ce{H2S} ice upper limits), and ice thicknesses larger ($\gtrsim$0.01 $\mu$m; \citealt{Dartois2018}) than our experimental conditions. Like for higher dilutions, entrapment efficiencies have been shown to increase with ice thickness \citep{Fayolle2011, Simon2019, Bergner2022, Simon2023, Kruczkiewicz2024}, though this dependency might break down for thicknesses $\gtrsim$50 ML \citep{Pesciotta2024}. Moreover, the deposition temperatures used in our entrapment experiments generate porous amorphous water ice matrices, whereas more representative cASW might trap volatiles more efficiently than porous counterparts \citep{Kruczkiewicz2024}.\footnote{Previously, \cite{Burke2010} suggested that more compact ASW should result in lower entrapment efficiencies due to a reduction in available trapping sites. This interpretation was based on earlier experiments by \cite{BarNun1988}, which observed lower entrapment efficiencies of hypervolatiles in binary mixtures with water ice as the deposition temperature increased. However, this decline is more likely dominated by the reduced surface residence time of hypervolatiles at higher deposition temperatures \citep{Zhou2024}, rather than due to the water ice morphology. Recently, \cite{Kruczkiewicz2024} deposited cASW ices at 10 K using well-collimated beams, and found that trapping capacities were actually higher for the compact ice structure.} These three factors (mixing ratios, thicknesses, and water ice morphology) thus point to real \ce{H2S} entrapment efficiencies being higher than the values measured here, further highlighting the shift in the \ce{H2S} sublimation front closer to the water snowline. At the same time, differences between the heating rates used in the laboratory and those occurring in astrophysical timescales could result in an overestimation of the measured entrapment efficiencies \citep{Cuppen2017}. Moreover, since sulfur atoms are heavier and less abundant than oxygen, the bulk of the \ce{H2S} ice has been predicted to form at slightly later timescales (by $\sim$1.4 A$_\text{V}$) than \ce{H2O} \citep{Goicoechea2021}. This differential formation could result in a concentration gradient for \ce{H2S} within the polar ice layer, meaning that a fraction of \ce{H2S} could exists in higher local concentrations than the estimated $\lesssim$1\%. In any case, the very efficient entrapment of \ce{H2S} in water, with measured efficiencies of $\gtrsim$75\% for \ce{H2S} concentrations even as high as $\sim$20\%, means that even in these scenarios, a significant portion of \ce{H2S} will remain entrapped in water. 

Some ice species, such as \ce{CO2} and potentially \ce{HCN}, exhibit a segregation behavior upon heating, where diffusion leads to the formation of pockets of pure ice instead of a homogeneous mixture (e.g., \citealt{Oberg2009, Boogert2015, Bergner2022}). For \ce{H2S}, segregation appears to be less energetically favorable than for the archetypal case of \ce{CO2}. Unlike \ce{CO2}, the \ce{H2S}$-$\ce{H2O} binding energies exceed that of \ce{H2S}$-$\ce{H2S}, and \ce{H2S} appears to wet the cASW surface effectively. However, the significantly stronger stabilization among \ce{H2O} molecules themselves compared to \ce{H2S}$-$\ce{H2O} interactions ($E_b$(\ce{H2O}$-$\ce{H2O}) $\sim$ 5600 K; \citealt{Wakelam2017}) may still drive some degree of segregation, as a system with separate water-rich and \ce{H2S}-rich pockets can be energetically more favorable than a fully mixed one. Similar energetic considerations are required in the kinetic Monte Carlo simulations performed by \cite{Oberg2009} to reproduce the segregation behavior they observed experimentally for \ce{CO2}:\ce{H2O} mixed ices. \footnote{\cite{Oberg2009} find that segregation in \ce{CO2}:\ce{H2O} ice mixtures is not precluded by \ce{CO2}$-$\ce{CO2} binding energies being lower than those of \ce{CO2}$-$\ce{H2O}. Rather, it might occur when a \ce{CO2} molecule from a water-dominated environment is swapped with a \ce{H2O} molecule from a \ce{CO2}-dominated environment. This process only requires that 0.5$\times$E$_b$(\ce{H2O}$−$\ce{H2O})$+$0.5$\times$E$_b$(\ce{CO2}$−$\ce{CO2}) is greater than E$_b$(\ce{H2O}$−$\ce{CO2}).} Consequently, in an astrophysical context, a fraction of \ce{H2S} within channels with access to the surface might desorb at the sublimation fronts defined by \ce{H2S}$-$\ce{H2S} and \ce{H2S}$-$\ce{H2O} interactions, depending on the extent of segregation.

Nonetheless, the primary sublimation front of \ce{H2S} is driven by its entrapment. This effectively shifts the \ce{H2S} sublimation front closer to the protostar by nearly threefold, fully retaining \ce{H2S} in the ice throughout and beyond the comet-formation zone ($>5$ au; \citealt{Mandt2015}), and across the region where other \ce{H2O}-rich bodies, such as icy asteroids, are formed. Consequently, \ce{H2S} may be incorporated into planetesimal cores formed beyond the water snowline ($\sim$1.9 au), which in turn may deliver it to terrestrial planets. In fact, Rosetta measurements of the \ce{H2S}/\ce{H2O} ratios in the coma of comet 67P remained remarkably constant over several months, strongly suggesting that \ce{H2S} desorbs from the comet nucleus along with \ce{H2O} \citep{Calmonte2016}. This supports the hypothesis that \ce{H2S} is incorporated into cometary cores mixed with \ce{H2O}, where it remains preserved beyond its sublimation temperature, until water ice desorption.

Additionally, solid-state \ce{H2S} can react with \ce{NH3} to form the ammonium salt \ce{NH4^{+}SH^{-}}---a likely major carrier of both nitrogen and sulfur in comets \citep{Altwegg2022, Slavicinska2024b}. Upon sublimation, this salt decomposes back into its neutral reactants, releasing \ce{H2S} (and \ce{NH3)} into the gas phase. The desorption temperature of the salt is nearly identical to that of water, meaning that the gaseous \ce{H2S} released from the decomposition of the salt will have a sublimation front similar to that of neutral \ce{H2S} entrapped in \ce{H2O}. Distinguishing the contribution from the neutral \ce{H2S} and the salt to the gas-phase distribution of \ce{H2S} in disks presents a challenge, but it could be an interesting avenue for future exploration. We emphasize that the neutral \ce{H2S} component within interstellar ices is likely significant, as the \ce{NH4^{+}SH^{-}} salt detected by the Rosetta mission is found in the comet's grains, while the ice predominantly contains \ce{H2S} in its neutral form \citep{Altwegg2022}.

\section{Conclusions} \label{sec:concl}

In this work, we provide a comprehensive characterization of the thermal sublimation dynamics of \ce{H2S} ice. We investigate its binding energies both dominated by interactions with other \ce{H2S} molecules (\ce{H2S}$-$\ce{H2S}) and with \ce{H2O} molecules (\ce{H2S}$-$\ce{H2O}), as well as the efficiency with which it is entrapped in water-rich ices. This information is used to estimate the different snowline positions of \ce{H2S} in the midplane of a representative T-Tauri protoplanetary disk. Our main findings are as follows:

\begin{enumerate}
    \item We derive $E_b=3159\pm46$ K for the \ce{H2S}$-$\ce{H2S} binding energy, with a mean preexponential factor of $\nu=(1.76\pm0.07)\times10^{16}$ ML s$^{-1}$. 
    
    \item For \ce{H2S}$-$\ce{H2O}, we find four sets of desorption parameters for four different \ce{H2S} coverages (0.6, 0.5, 0.3, and 0.2 ML). The mean binding energy and preexponential factor are $E_b=3392\pm 56$ K and $\nu=(1.7\pm 0.1)\times10^{16}$ s$^{-1}$.

    \item Theoretical \ce{H2S} binding energies generally underpredict the experimental values derived in this work. We propose this mismatch is due to a limitation in the computations to account for adsorbate diffusion causing a preference for occupying deeper surface biding sites. In interstellar and protoplanetary conditions, diffusion plays an important role in a molecule's sublimation dynamics, and therefore our experimentally-derived binding energies are expected to be more representative values.

    \item \ce{H2S} is very efficiently entrapped in \ce{H2O} ice, with efficiencies of $\sim75 - 85 \%$ for \ce{H2O}:\ce{H2S} mixing ratios of 1:$\sim5-17$ (more diluted cases yielding more efficient entrapment). In comparison, \ce{^{13}CO2} is much less efficiently entrapped, with an eficiency of $\sim$52\% for a \ce{H2O}:\ce{^{13}CO2} ratio of 1:14. We suggest this might be due to the hydrogen bonding networks between \ce{H2S} and \ce{H2O}, which are stronger intermolecular interactions that the induced dipoles between \ce{^{13}CO2} and \ce{H2O}. Pore effects could also play a role in entrapping \ce{H2S} more efficiently than \ce{^{13}CO2}.

    \item Our measured \ce{H2S}$-$\ce{H2S} and \ce{H2S}$-$\ce{H2O} binding energies yield \ce{H2S} sublimation temperatures at $\sim$64 K and $\sim$69 K, respectively. This corresponds to a radial distance of $\sim$6.3 au and $\sim$5.6 au in the midplane of a representative T-Tauri disk. 

    \item The vast majority of \ce{H2S} will remain entrapped in the ice until water crystallizes at radii close to the \ce{H2O} snowline, shifting its sublimation front inward by nearly a factor of three. As a result, most of the \ce{H2S} will remain in ices throughout the region where water-rich icy planetesimals form.

\end{enumerate}
\begin{acknowledgements}
J.C.S. acknowledges support from the Danish National Research Foundation through the Center of Excellence “InterCat” (Grant agreement no.: DNRF150) and the Leiden University Fund / Fonds Van Trigt (Grant reference no.: W232310-1-055). K.I.\"{O}. acknowledges an award from the Simons Foundation (\#321183FY19).
\end{acknowledgements}


\appendix

\section{Submonolayer coverage estimation}
\label{Appendix:subML_calib}
\begin{figure}[htb!]\centering
\includegraphics[scale=0.35]{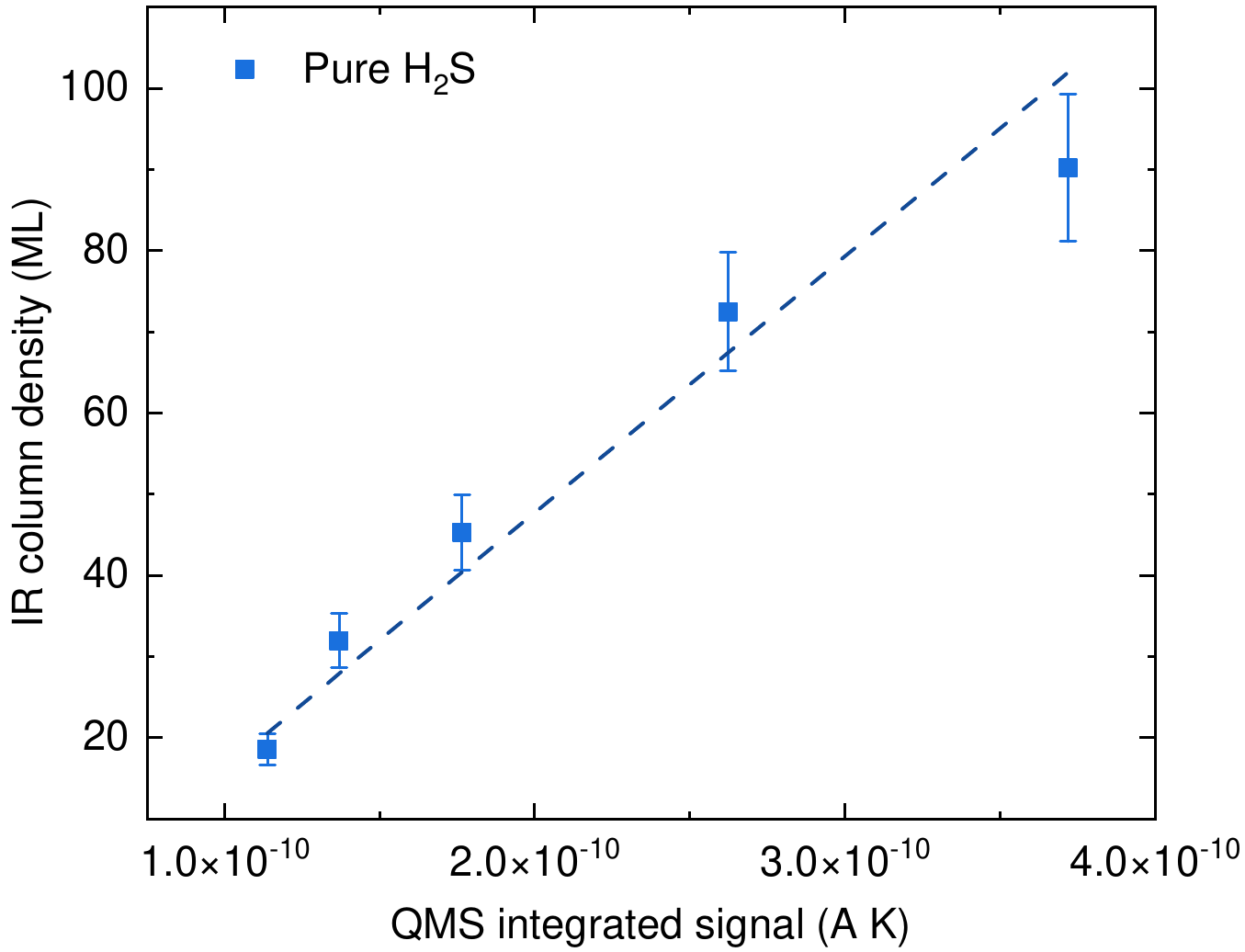}
\caption{\ce{H2S} column densities derived from the integrated absorbance of its S-H stretching modes at 15 K for each multilayer ice thickness as a function of its corresponding integrated desorption feature as measured by the QMS. The dashed blue line shows the linear fit to the points from which the QMS-to-column density scaling factor was derived.}
\label{fig:qms_ir_convert}
\end{figure}

The submonolayer \ce{H2S} coverages were estimated based on a QMS-to-column density scaling factor, $f$, derived from the multilayer \ce{H2S} experiments. Figure \ref{fig:qms_ir_convert} shows the correlation between the \ce{H2S} ice column density obtained from the integrated absorbance of its S-H stretching feature (see Equation \ref{eq:N_RAIRS}) and the integrated signal of the \ce{H2S} desorption feature as measured by the QMS for m/z = 34.

A linear fit to the plot gives the conversion factor $f=(3.2\pm0.4)\times10^{26}$ A$^{-1}$ K$^{-1}$. In the low coverage experiments, where \ce{H2S} was deposited on top of cASW, the \ce{H2S} ices with approximately 1.6 ML and 1.2 ML (determined from their infrared absorbance bands) produced S-H stretching features detectable above our instrumental noise. By comparing the measured and predicted \ce{H2S} coverages in these experiments, we estimate that the submonolayer coverages might be underpredicted by up to a factor of 2 using our scaling method. However, this discrepancy does not impact our analysis, as the higher coverages (by a factor of two) yield binding energy distributions with mean E$_b$ values differing by less than 2 K from those derived for the predicted coverages---well within their FWHMs.

\section{Pure \ce{H2S} ice infrared features vs temperature}
\label{Appendix:H2S_multiL_IR_TPD}

\begin{figure*}[htb!]\centering
\includegraphics[scale=0.45]{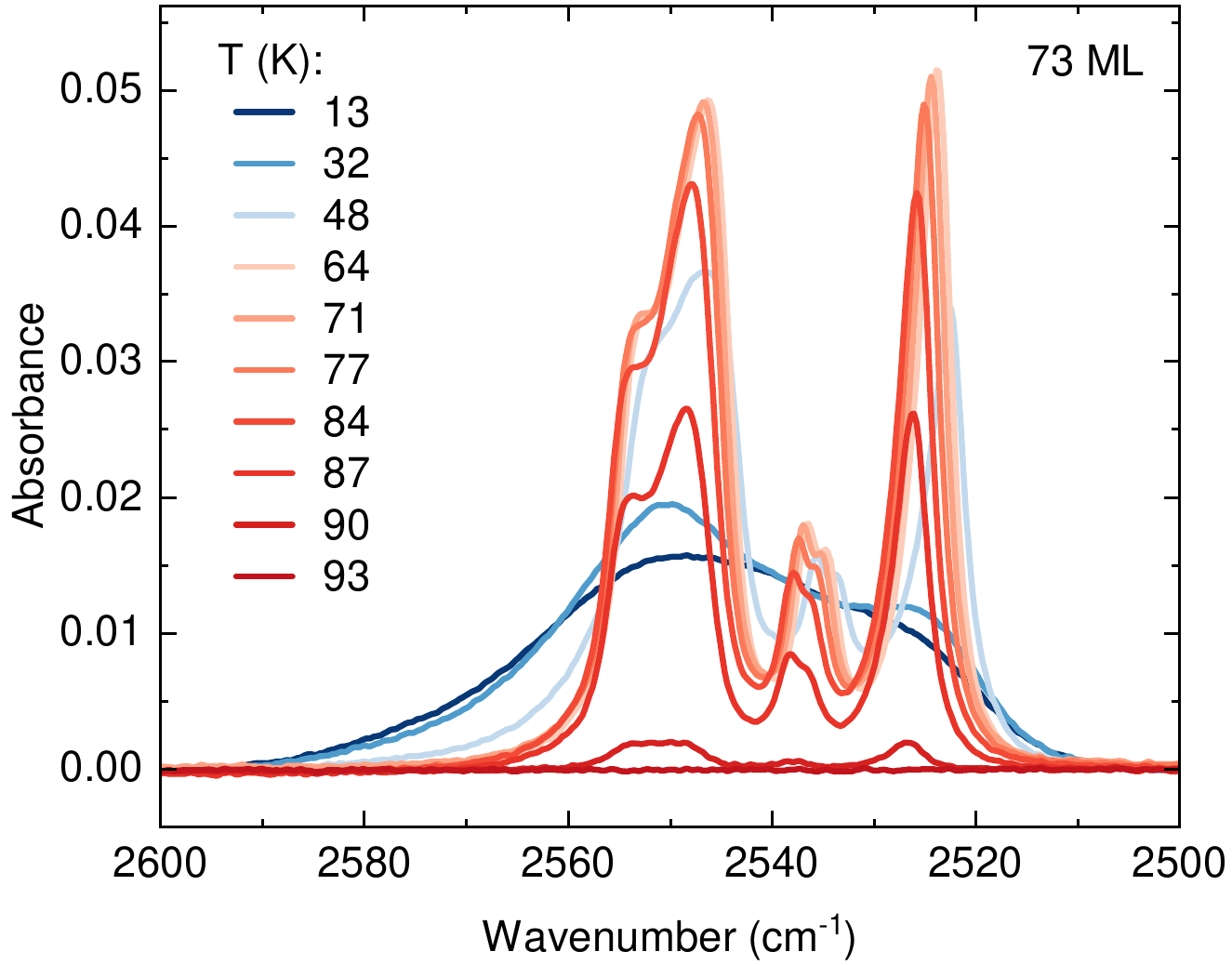}
\caption{Infrared spectra taken during the TPD experiment of the 73 ML \ce{H2S} ice, centered on the frequency range of its S-H stretching modes. For clarity, only a subset of spectra collected at the relevant temperatures are shown.}
\label{fig:multiL_H2S_IR_TPD}
\end{figure*}

Figure \ref{fig:multiL_H2S_IR_TPD} shows the IR spectra recorded during the TPD experiment of a multilayer \ce{H2S} ice with an initial coverage of 73 ML. The splitting between its symmetric ($\nu_1$) and antisymmetric ($\nu_3$) S-H stretching modes signals that ice crystallization starts to occur at temperatures as low as $\sim$30 K. By $\sim$65 K, the transition to phase III crystalline \ce{H2S} is nearly complete. However, from $\sim$65 K up to the point of complete desorption, a continuous blueshift in the S-H stretching features is observed, suggesting ongoing structural reorganization within the ice. While higher-energy crystalline phases of \ce{H2S} exist (phase II transitioning at $\sim$100 K and phase I transitioning at $\sim$125 K, see \citealt{Fathe2006} and references therein), these transitions are observed under ambient pressure and occur above the sublimation temperature of \ce{H2S} in UHV conditions. They are therefore not relevant to our experiments.

\section{Arrhenius plots}
\label{Appendix:multiL_arrhenius}

\begin{figure*}[htb!]\centering
\includegraphics[scale=0.3]{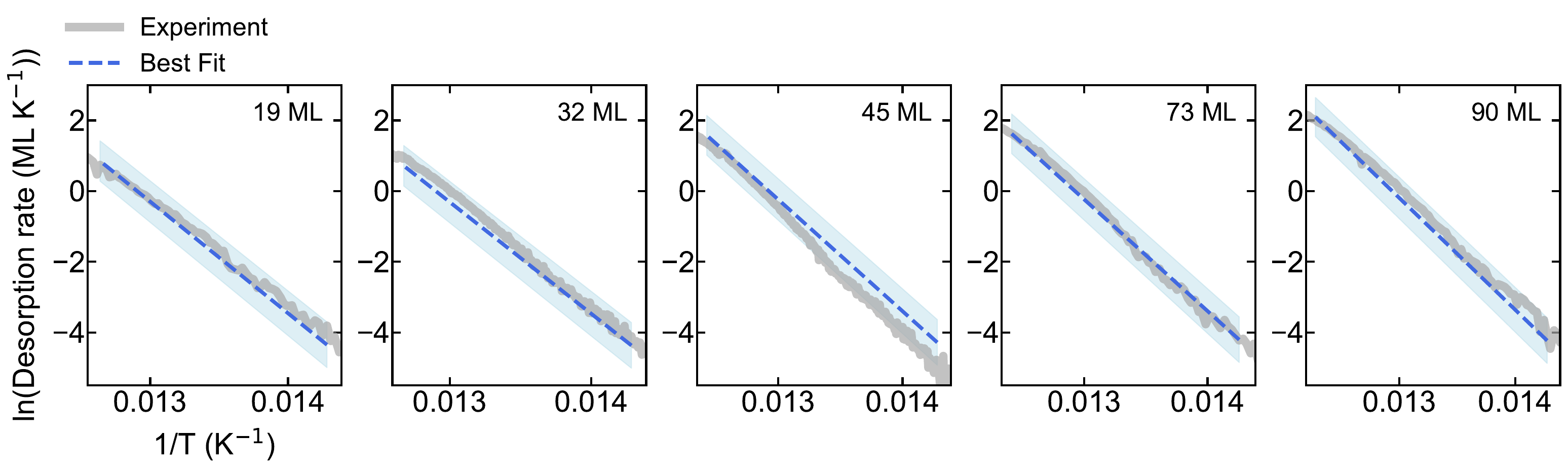}
\caption{Arrhenius plots of the multilayer \ce{H2S} TPD curves used to derive the \ce{H2S}$-$\ce{H2S} binding energy. The experimental data is shown in gray, while the linear fits to the plots, performed simultaneously for the five ice thicknesses, are shown by the blue dashed lines. The shaded blue region indicates the 1$\sigma$ uncertainty. The fit is performed for the temperature range where the original curve follows an exponential trend (see section \ref{sec:BE_multiL}).}
\label{fig:TPD_multi_arrhenius}
\end{figure*}

Figure \ref{fig:TPD_multi_arrhenius} shows the Arrhenius plots of the TPD experiments of all multilayer \ce{H2S} ice coverages. The desorption rate data (gray) for all five thicknesses was fit simultaneously following a Monte-Carlo sampling approach with 10,000 trials. The mean best-fit model corresponds to an \ce{H2S}-\ce{H2S} binding energy of E$_b$ = $3159\pm46$ K (blue dashed lines).

\section{Multilayer \ce{H2S} ice fits with a temperature-dependent $\nu_\text{TST}$}
\label{Appendix:T_dependent_nuTST}

\begin{figure*}[htb!]\centering
\includegraphics[scale=0.3]{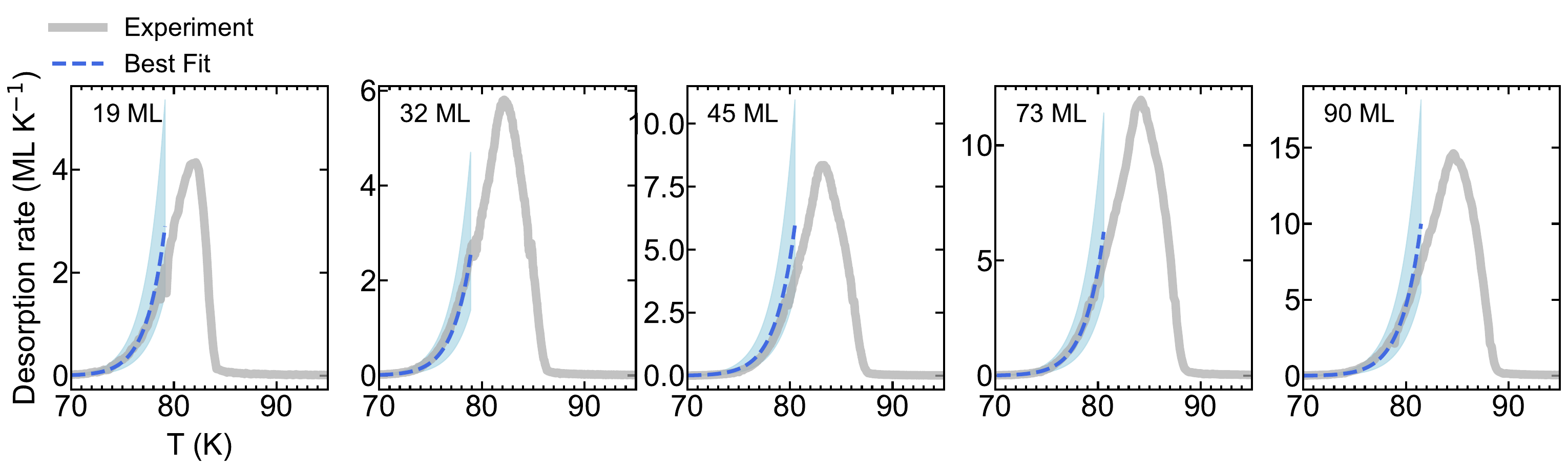}
\caption{Same as Figure \ref{fig:TPD_multi_leadingedge}, but with fits performed using a temperature-dependent $\nu_\text{TST}$.}
\label{fig:multiL_fits_all_nuTST_Tdependent}
\end{figure*}

\begin{figure*}[htb!]\centering
\includegraphics[scale=0.3]{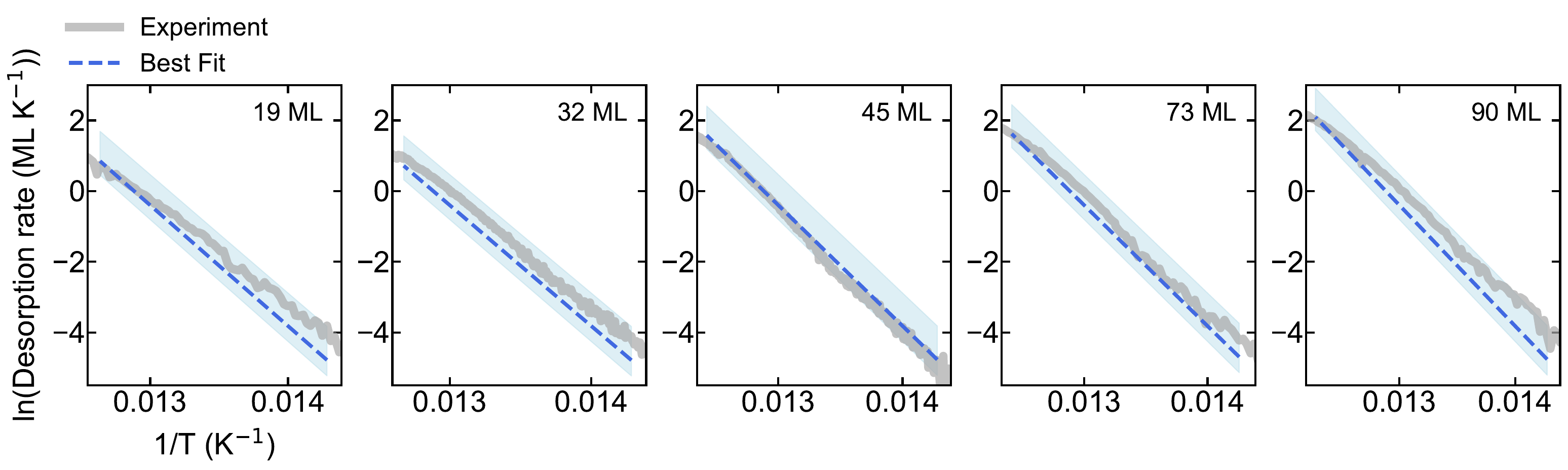}
\caption{Same as Figure \ref{fig:TPD_multi_arrhenius}, but with fits performed using a temperature-dependent $\nu_\text{TST}$.}
\label{fig:multiL_fits_all_nuTST_arrhenius_Tdependent}
\end{figure*}

Figures \ref{fig:multiL_fits_all_nuTST_Tdependent} and \ref{fig:multiL_fits_all_nuTST_arrhenius_Tdependent} show the fits to the log-transformed multilayer TPD curves of \ce{H2S} with a temperature-dependent $\nu_\text{TST}$. The former presents the data as the original TPD profile, while the latter shows the corresponding Arrhenius plots. This approach yields E$_b$ = $3141\pm49$ K (blue dashed lines).

\section{Submonolayer fit with a single energy component}
\label{Appendix:subML_oneE}

\begin{figure}[htb!]\centering
\includegraphics[scale=0.3]{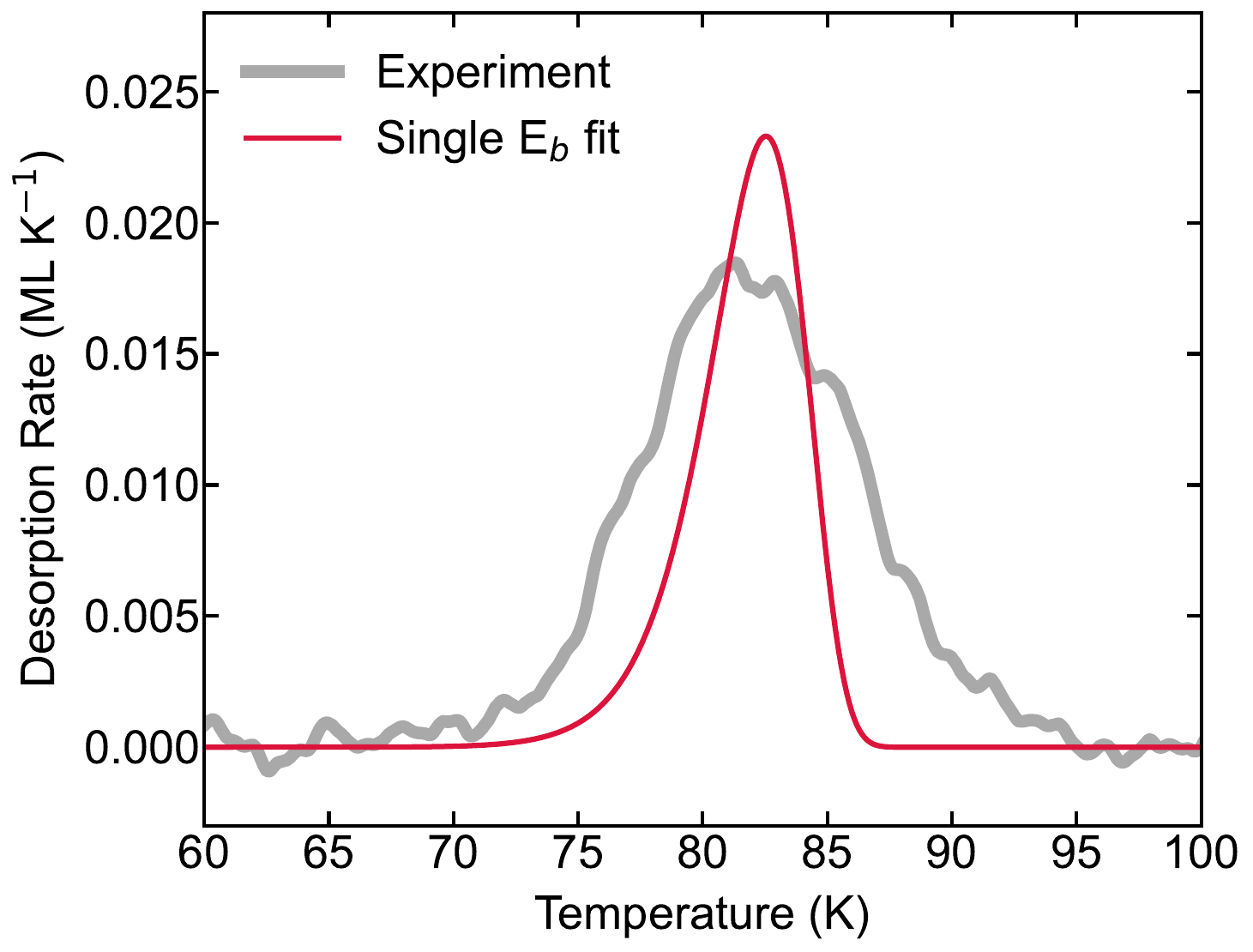}
\caption{TPD curve of the 0.2 ML \ce{H2S} ice experiment deposited on top of a cASW substrate. The experimental data is show in gray, while the attempted fit to the data with a first-order Polanyi-Wigner curve with only one temperature component is shown in red.}
\label{fig:02ML_fit_single}
\end{figure}

Figure \ref{fig:02ML_fit_single} illustrates an attempted fit of the 0.2 ML \ce{H2S} ice using the Polanyi-Wigner equation (eq. \ref{eq:Polanyi}) with a single temperature component. The fit fails to capture the experimental data, highlighting the presence of a distribution of binding energies caused by the inherent inhomogeneity of the cASW substrate.

\section{QMS-TPD results for mixed \ce{H2O}:\ce{H2S} ices}
\label{Appendix:TPD_mixed_ices}

\begin{figure*}[htb!]\centering
\includegraphics[scale=0.6]{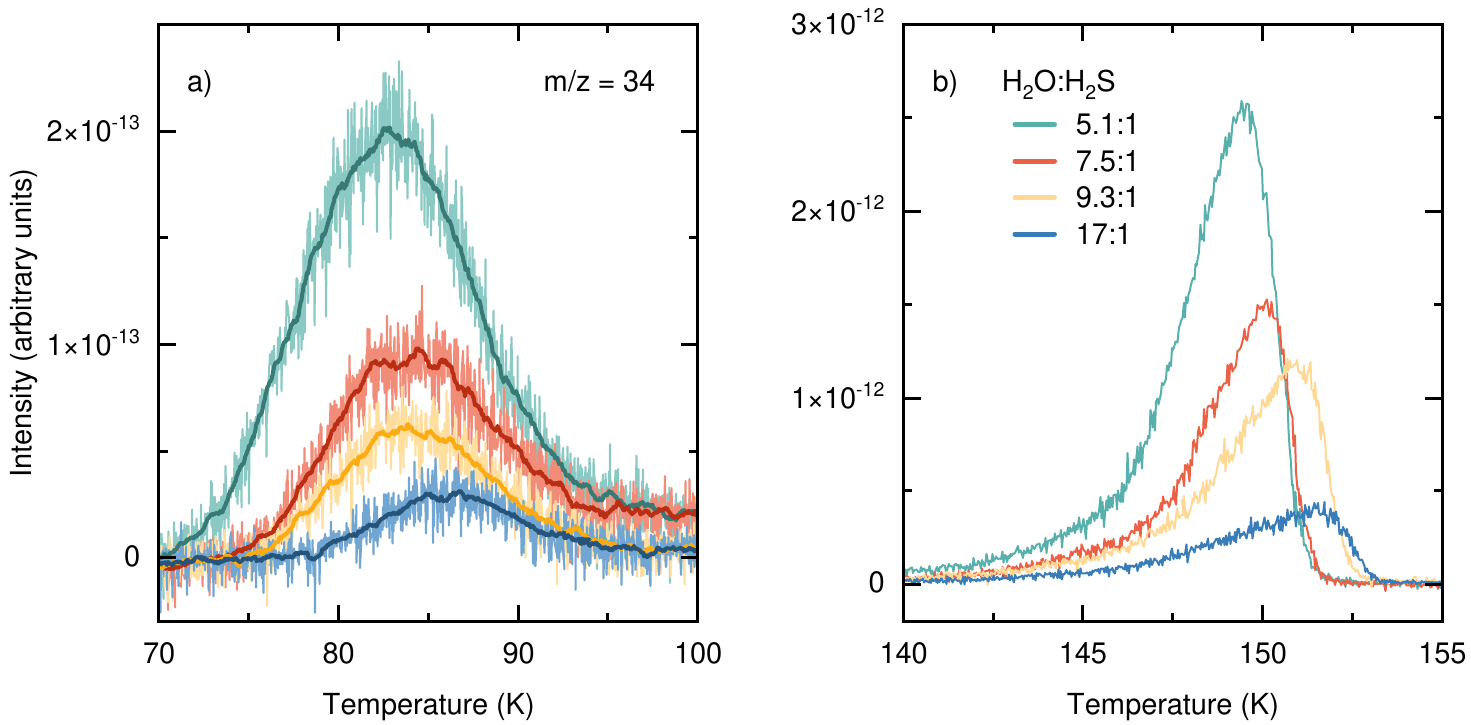}
\caption{TPD data measured by the QMS for m/z = 34 in the experiments with mixed \ce{H2O}:\ce{H2S} ices. Panel a shows the monolayer desorption of \ce{H2S} molecules with access to the surface, and panel b depicts the molecular volcano feature due to \ce{H2S} entrapped in the \ce{H2O} ice matrix.}
\label{fig:entrapped_all_TPDs}
\end{figure*}

Figure \ref{fig:entrapped_all_TPDs} presents the QMS data collected for m/z = 34 during the TPD experiments with mixed \ce{H2O}:\ce{H2S} ices. In all cases, the first desorption feature (panel a) corresponds to monolayer desorption characterized by the \ce{H2S}$-$\ce{H2O} binding energies. This is evinced by the peak desorption temperatures, which increase with decreasing coverages, and by the desorption profiles consistent with first-order desorption kinetics for all mixing ratios. The second desorption feature (panel b) corresponds to the molecular volcano.

%
%

   \bibliographystyle{aa} 
   \bibliography{mybib} 

\end{document}